\documentclass[reqno]{amsart}

\usepackage{lineno,hyperref}

\usepackage{graphicx}          

\usepackage{booktabs} 
\usepackage{IEEEtrantools}
\usepackage{amssymb,latexsym,amsfonts,amsmath}
\usepackage{graphicx}
\usepackage{dsfont}
\usepackage{ wasysym }
\usepackage{multirow}

\usepackage{tikz}
\usetikzlibrary{calc,shapes,arrows}

\usepackage{mdframed,lipsum}

\usepackage{mathrsfs}

\usepackage{algorithm}
\usepackage{algorithmic}
\usepackage{url}

\usepackage[many]{tcolorbox}
\usetikzlibrary{calc}
\tcbuselibrary{skins}

\newtcolorbox{resp}[1][]{%
	enhanced jigsaw,%
	colback=gray!5!white,%
	colframe=gray!80!black,%
	size=small,%
	boxrule=1pt,%
	halign title=flush center,%
	coltitle=black,%
	breakable,%
	drop shadow=black!50!white,%
	attach boxed title to top left={xshift=1cm,yshift=-\tcboxedtitleheight/2,yshifttext=-\tcboxedtitleheight/2},%
	minipage boxed title=3cm,%
	boxed title style={%
		colback=white,%
		size=fbox,%
		boxrule=1pt,%
		boxsep=2pt,%
		underlay={%
			\coordinate (dotA) at ($(interior.west) + (-0.5pt,0)$);
			\coordinate (dotB) at ($(interior.east) + (0.5pt,0)$);
			\begin{scope}[gray!80!black]
				\fill (dotA) circle (2pt);
				\fill (dotB) circle (2pt);
			\end{scope}
		}%
	},%
	#1%
}

\usepackage{xspace}

\newcommand{\R}{{\mathbb{R}}}

\newcommand{\N}{{\mathbb{N}}}

\newcommand{\Let}{:=}

\newtheorem{theorem}{Theorem}[section]
\newtheorem{lemma}[theorem]{Lemma}
\newtheorem{problem}[theorem]{Problem}

\newtheorem{corollary}[theorem]{Corollary}

\newtheorem{definition}[theorem]{Definition}
\newtheorem{example}[theorem]{Example}

\newtheorem{remark}[theorem]{Remark}
\newtheorem{assumption}[theorem]{Assumption}

\renewcommand{\emptyset}{{\varnothing}}

\begin{document}

\title[Compositional Synthesis of Barrier Certificates for Networks of Stochastic Systems]{Compositional Synthesis of Control Barrier Certificates for Networks of Stochastic Systems against $\omega$-Regular Specifications}

\author{Mahathi Anand$^{1}$}
\author{Abolfazl Lavaei$^{2}$}
\author{Majid Zamani$^{3,1}$}
\address{$^1$Department of Computer Science, LMU Munich, Germany}
\email{mahathi.anand@lmu.de}
\address{$^2$Institute for Dynamic Systems and Control, ETH Zurich, Switzerland}
\email{alavaei@ethz.ch}
\address{$^3$Department of Computer Science, University of Colorado Boulder, USA.}
\email{majid.zamani@colorado.edu}

\begin{abstract}
	This paper is concerned with a compositional scheme for the construction of control barrier certificates for interconnected discrete-time stochastic systems. The main objective is to synthesize switching control policies against $\omega$-regular properties that can be described by accepting languages of deterministic Streett automata (DSA) along with providing probabilistic guarantees for the satisfaction of such specifications. The proposed framework leverages
	the interconnection topology and a notion of so-called \emph{control sub-barrier certificates} of subsystems, which are used to compositionally construct control barrier certificates of interconnected systems by imposing some dissipativity-type compositionality conditions. We propose a systematic approach to decompose high-level $\omega$-regular specifications into simpler tasks by utilizing the automata corresponding to the specifications. In addition, we formulate an alternating direction method of multipliers (ADMM) optimization problem in order to obtain suitable control sub-barrier certificates of subsystems while satisfying compositionality conditions. For systems with polynomial dynamics, we provide a sum-of-squares (SOS) optimization problem for the computation of control sub-barrier certificates and local control policies of subsystems. Finally, we demonstrate the effectiveness of our proposed approaches by applying them to a physical case study.
\end{abstract}

\maketitle

\section{Introduction}

Formal verification and synthesis of complex dynamical systems against high-level logic specifications, \emph{e.g.,} those expressed as
linear temporal logic (LTL) formulae~\cite{pnueli1977temporal}, have gained remarkable attentions over the past few years~\cite{Tabuada.2009}. Such problems are particularly challenging when dealing with high-dimensional continuous-state systems with stochasticity inside the models, especially, in real-world safety-critical scenarios such as traffic networks, power grids, etc.   

Verification and synthesis of large-scale stochastic systems have been discussed in the relevant literature mainly based on abstraction-based techniques. Some existing results include providing probabilistic guarantees for discrete-time stochastic hybrid systems against safety and reachability properties~\cite{Abate2008}, game-based abstractions for verification and controller synthesis of hybrid automata enforcing reachability specifications~\cite{kattenbelt_game-based_2010,hahn_game-based_2011}, and symbolic controller synthesis for Markov decision processes against B{\"u}chi specifications~\cite{majumdar_symbolic_2020}. These techniques, however, rely on state-space discretization and therefore suffer severely from the \emph{curse of dimensionality}: the computational complexity increases exponentially with respect to the dimension of the state space. Hence, these approaches are not applicable to large-scale complex systems. This issue has been partly mitigated by  utilizing adaptive sequential gridding algorithms~\cite{EsmaeilZadehSoudjani.2013} or input-set abstractions for incrementally stable stochastic control systems~\cite{Zamani.2017}. Another promising alternative to alleviate the state-explosion problem is to consider the large-scale complex system as an interconnection of smaller subsystems and construct finite abstractions of the interconnected system from abstractions of subsystems via compositional techniques~\cite{SAM17,lavaei2018ADHSJ,lavaei2018CDCJ,lavaei2019HSCC_J,lavaei2019NAHS,lavaei2017HSCC,lavaei2020NAHSJ,lavaei2019Thesis,AmyJournal2020,AmyIFAC2020}. 

More recently, discretization-free approaches via control
barrier certificates have been proposed for the verification and
synthesis of stochastic systems. Existing results include verification and synthesis of continuous and hybrid stochastic systems over infinite-time horizons~\cite{prajna2007,Huang.2017,Wisniewski.2018}, safety verification of discrete-time stochastic systems~\cite{ahmadi_verification_2018}, verification and synthesis of discrete-time stochastic control systems against LTL specifications over finite-time horizons~\cite{jagtap2018temporal,jagtap_formal_2019}, compositional construction of control barrier certificates for stochastic switched
systems via max small-gain conditions~\cite{nejati_lcss}, and recently, control barrier functions for stochastic systems in the presence of process and measurement noise~\cite{clark2020control}.  

The proposed results in the above-mentioned literature require control barrier certificates to have a certain parametric form, such as polynomial, in order to search for their unknown coefficients under some mild assumptions. Although searching for those coefficients might be possible for systems with lower dimensions, it will become increasingly difficult and computationally intractable for large-scale complex systems. In order to alleviate this issue, one promising solution is to develop a compositional approach by considering the large-scale stochastic system as an interconnection of lower-dimensional subsystems and searching for so-called \emph{control sub-barrier certificates} for subsystems along with the corresponding local control policies. One can then utilize  control sub-barrier certificates of subsystems to compositionally construct a control barrier certificate of the complex monolithic system.

In this paper, we propose a compositional technique based on dissipativity approaches for the construction of control barrier certificates for interconnected discrete-time stochastic systems. The proposed compositional condition may leverage the structure of the interconnection topology together with dissipativity-type properties of subsystems and be potentially fulfilled independently of the number or gains of the subsystems. The main goal is to synthesize switching control policies against $\omega$-regular properties that can be described by accepting languages of deterministic Streett automata (DSA). We provide a systematic approach to decompose high-level $\omega$-regular specifications into simpler tasks by employing the automata corresponding to the specifications. Control sub-barrier certificates are obtained for these tasks along with the corresponding local control policies. Correspondingly, a switching control policy that ensures the satisfaction of the specification with some probability is synthesized.

Conventionally, in order to satisfy the compositionality condition, the required parameters for finding suitable control sub-barrier certificates are pre-selected and the compositionality condition is checked a posteriori. While this method can provide tractable results in certain scenarios for systems with specific interconnection structures, it is not particularly useful in large-scale networks where structural properties of interconnected systems are not apparent. Besides, since control sub-barrier certificates are not optimized with respect to the compositionality condition, obtained results can be conservative. In order to provide scalable, less-conservative results, we employ a distributed optimization method based on an alternating direction method of multipliers (ADMM) algorithm which allows us to break down a large optimization problem into several smaller sub-problems which can be easier to handle. The solution to the optimization problem provides us with suitable control sub-barrier certificates along with local control policies, allowing computation of control barrier certificates for the interconnected system. For systems with polynomial dynamics, we show that ADMM algorithm can be utilized in conjunction with sum-of-squares (SOS) optimization in order to obtain control sub-barrier certificates and corresponding local control policies. We demonstrate the effectiveness of our proposed results by applying them to a room temperature network in a circular building containing $300$ rooms.

{\bf Related Literature.} Compositional construction of control barrier certificates for discrete-time stochastic systems enforcing simple safety specifications is recently proposed in~\cite{mahathi_IFAC2020} and then extended in \cite{anand2021smallgain} for a larger class of specifications that can be admitted by accepting languages of deterministic finite automata (DFA). Both works use a different compositionality scheme, namely, \emph{$\max$ small-gain} conditions. Our approach here differs from the one in~\cite{anand2021smallgain} in several main directions. First and foremost, our proposed compositionality approach here is potentially less conservative than the one presented in~\cite{anand2021smallgain} since the dissipativity-type reasoning proposed in this work can leverage the structure of the interconnection topology and may not require any constraint on the number or gains of the subsystems for some specific interconnection structures, \emph{e.g.,} skew-symmetric (cf. Remark~\ref{compositionality remark}). Second, the small-gain approach in~\cite{anand2021smallgain} requires the satisfaction of a circular compositionality condition (see~\cite[condition 4.4]{anand2021smallgain}), which is difficult to be checked systematically for large interconnected systems. In contrast, our compositionality condition here is a simple linear matrix inequality (LMI), which is easy to check. Third, the provided results in~\cite{anand2021smallgain} asks an additional condition (\emph{i.e.,}~\cite[condition (3.1)]{anand2021smallgain}) which is required for the satisfaction of \emph{small-gain} type conditions, while we do not need such an assumption in our setting. Fourth, while~\cite{anand2021smallgain} provides probabilistic guarantees for finite-time horizons, we generalize those results to \emph{infinite-time} horizons. Consequently, we enlarge the class of specifications here to $\omega$-regular properties. As our last contribution, we propose here a distributed optimization method based on an ADMM algorithm which allows us to break down a large optimization problem into several smaller sub-problems. In particular, the solution to the optimization problem provides suitable control sub-barrier certificates together with local control policies satisfying the main compositionality condition and, hence, allowing us to compute control barrier certificates for the interconnected system efficiently. In comparison, one needs to first compute control sub-barrier certificates in~\cite{anand2021smallgain} for each subsystem and then check the compositionality condition a-posteriori. In the case that the compositionality condition in~\cite{anand2021smallgain} is not fulfilled, one requires to re-design the control sub-barrier certificates from scratch again.

Compositional construction of control barrier functions via $\max$ small-gain conditions for interconnected control systems is also presented in~\cite{jagtap2020compositional}. Our proposed approach differs from the one in~\cite{jagtap2020compositional} in six directions, four of which are common with the comparisons with the small-gain conditions that we discussed above for~\cite{mahathi_IFAC2020}. As the fifth distinction, the proposed results in~\cite{jagtap2020compositional} are presented for non-stochastic settings while our work deals with stochastic ones. As our last contribution, the results in~\cite{jagtap2020compositional} are provided for specifications described by deterministic co-B\"uchi automata (a subset of $\omega$-regular specifications), whereas we deal with general $\omega$-regular specifications in this paper. Lastly, compositional construction of control barrier functions for large-scale stochastic systems has also been presented in \cite{nejati2020compositional,AmyIFAC12020} but for continuous-time stochastic systems using a different compositional scheme, namely, sum-type small-gain conditions. Unfortunately, those conditions are formulated in terms of “almost” linear gains and require subsystems to have a (nearly) linear behavior, making it much more conservative than our proposed approach here.

\section{Discrete-Time Stochastic Control Systems}

\subsection{Preliminaries}

We consider the probability space $(\Omega, \mathcal{F}_\Omega, \mathbb{P}_\Omega)$, where $\Omega$ is the sample space, $\mathcal{F}_\Omega$ is a sigma-algebra on $\Omega$ consisting subsets of $\Omega$ as events, and $\mathbb{P}_\Omega$ is the probability measure that assigns probability to those events. Random variables in this paper are assumed to be measurable functions of the form $X:(\Omega,\mathcal{F}_\Omega) \rightarrow (S_X,\mathcal{F}_X)$. Any random variable $X$ induces a probability measure on $(S_X,\mathcal{F}_X)$ as $Prob\{A\} = \mathbb{P}_\Omega\{X^{-1}(A)\}$ for any $A \in \mathcal{F}_X.$ The topological space $S$ is a Borel space if it is homeomorphic to a Borel subset of a Polish space, \textit{i.e.}, a separable and completely metrizable space. The Borel sigma-algebra generated from Borel space $S$ is denoted by $\mathcal{B}(S)$ and the map $f: S \rightarrow Y$ is measurable whenever it is Borel measurable.

\subsection{Notations}

We use $\mathbb{R},\mathbb{R}_{>0}$, and $\mathbb{R}_{\geq 0}$ to denote the set of real, positive and non-negative real numbers, respectively, while $\mathbb{R}^n$ represents a real space of the dimension $n$. The set of non-negative integers and positive integers are denoted by $\mathbb{N} := \{0,1,\ldots\}$ and $\mathbb{N}_{\geq 1}=\{1,2,\ldots\}$, respectively. Given $N$ vectors $x_i \in \mathbb{R}^{n_i}$, we use $x=[x_1;\ldots;x_N]$ to denote the corresponding column vector of the dimension $\sum_i n_i$. We denote by $\mathsf{diag}(a_1,\ldots,a_N)$ a diagonal matrix in $\R^{N\times{N}}$ with diagonal entries $a_1,\ldots,a_N$. For a vector $x\in\mathbb{R}^{n}$, $\Vert x\Vert$ denotes the infinity norm of $x$. For a matrix $A \in \mathbb{R}^{n\times n}$, $\|A\|_F$ denotes the Frobenius norm of $A$. The identity matrix in $\mathbb R^{n\times{n}}$ is denoted by $\mathds{I}_n$. Given functions $f_i:X_i\rightarrow Y_i$, for any $i\in\{1,\ldots,N\}$, their Cartesian product $\prod_{i=1}^{N}f_i:\prod_{i=1}^{N}X_i\rightarrow\prod_{i=1}^{N}Y_i$ is defined as $(\prod_{i=1}^{N}f_i)(x_1,\ldots,x_N)=[f_1(x_1);\ldots;f_N(x_N)]$. To denote a set of vectors $\{x_1,\ldots,x_N\}$, we use the notation $x_{1:N}$.  For a set $S$, $|S|$ denotes its cardinality and $\emptyset$ denotes the empty set. The complement of the set $P$ with respect to $S$ is represented by $S\backslash P= \{x \, | \, x \in S, x \notin P\}$. The power set of $S$ is the set of all subsets of $S$ and is denoted by $2^{S}$. We employ $\top$ and $\bot$ to represent logical \textit{true} and \textit{false}, respectively. 

\subsection{Discrete-Time Stochastic Control Systems} \label{dtscs}

In this paper, we consider discrete-time stochastic control systems as formalized in the following definition.

\begin{definition} \label{def-dtSCS}
	A \emph{discrete-time stochastic control system} (dt-SCS) is a tuple 
	\begin{equation}\label{eq:dt-SCS}
	\mathfrak{S}=(X,U,W,\varsigma,f,Y,h),
	\end{equation}
	where,
	\begin{itemize}
		\item
		$X\subseteq \mathbb R^n$ is a Borel space as the state space of the
		system. The tuple $(X,\mathcal{B}(X))$ is the measurable state space where $\mathcal{B}(X)$ denotes the Borel sigma-algebra on the state space;
		\item $U\subseteq \mathbb R^m$ and $W\subseteq \mathbb R^p$ are Borel spaces as \emph{external} and \emph{internal} input spaces of the system; 
		\item
		$\varsigma:=\{\varsigma(k):\Omega\rightarrow \mathcal V_{\varsigma},\,\,k\in\N\}$ is a sequence of independent and identically distributed
		(i.i.d.) random variables from a sample space $\Omega$ to the
		measurable space
		$(\mathcal V_\varsigma, \mathcal F_\varsigma)$;	 
		\item
		$f:X \times U \times W \times\mathcal V_\varsigma\rightarrow X$ is a measurable function that
		characterizes the state evolution of $\mathfrak{S}$;
		\item 
		$Y\subseteq \mathbb{R}^{r}$ is a Borel space as \emph{external} output space of the system;
		\item 
		$h: X \rightarrow Y$ is a measurable function that maps a state $x \in X$ to its internal output $y=h(x)$.
	\end{itemize}
\end{definition}

We associate sets $\mathcal{U}$ and $\mathcal{W}$ to respectively sets $U$ and $W$ as collections of external and internal input sequences $\{\nu(k):\Omega\rightarrow U,\,\,k\in\mathbb N\}$ and $\{w(k):\Omega\rightarrow W,\,\,k\in\mathbb N\}$. Both $\nu(k)$ and $w(k)$ are independent of the random variable $\varsigma(l)$ for all $k,l \in \mathbb{N}$ and $l \ge k$. 
The state evolution of dt-SCS $\mathfrak{S}$ for a given initial state $x(0)\in
X$, and input sequences $\{\nu(k):\Omega\rightarrow U,\,\,k\in\mathbb N\}$ and $\{w(k):\Omega\rightarrow W,\,\,k\in\mathbb N\}$ is
characterized by:
\begin{equation}\label{Eq_1a}
\mathfrak S:\left\{\hspace{-0.5mm}\begin{array}{l}x(k+1)=f(x(k),\nu(k),w(k),\varsigma(k)),\\
y(k)=h(x(k)),\\
\end{array}\right.
\end{equation} 
for any $k\in\mathbb N$. 

For a given initial state $ a \in X$, $\nu(\cdot) \in \mathcal{U}$ and $w(\cdot) \in \mathcal{W}$, a random sequence $x^{a\nu w}:\Omega \times\mathbb N \rightarrow X$ denotes the solution process of $\mathfrak{S}$ under the influence of the internal input $\nu$, the external input $w$, and started from the initial state $a$.

The control of dt-SCS $\mathfrak{S}$ in ~\eqref{eq:dt-SCS} is enforced by history-dependent policies given by $\varpi = (\varpi_0,\varpi_1,\ldots)$ with $\varpi_i : \mathcal{D}_i \rightarrow U$, where $\mathcal{D}_i$ is the set of all $i$-histories $d_i$ that can be defined as $d_i := (x(0),\nu(0)$
$,x(1),\nu(1),\ldots,x(i-1),\nu(i-1),x(i))$. We consider a subclass of these policies called as stationary policies where $\varpi=(\nu,\nu,\ldots, \nu): X \rightarrow U$. In this case, the mapping at any time $i$ only depends on the current state $x(i)$ and is time-invariant.

The main focus of the paper is on the control of large-scale systems without internal inputs and outputs which can be regarded as a composition of smaller subsystems with internal inputs and outputs. Such a large-scale dt-SCS can be represented by the tuple $\mathfrak{S}=(X,U,\varsigma,f)$ 

where $f: X \times U \times \mathcal V_\varsigma\rightarrow X$. In this case, equation \eqref{Eq_1a} is reduced to
\begin{equation}\label{Eq_2a}
\mathfrak S: x(k+1)=f(x(k),\nu(k),\varsigma(k)),
\quad k\in\mathbb N.
\end{equation}
Note that although there is an internal output in the definition of dt-SCS in~\eqref{Eq_1a}, the full state information is assumed to be available for the large-scale system (\emph{i.e.,} its output map is identity) for the sake of controller synthesis. More precisely, the role of the internal output in~\eqref{Eq_1a} is mainly for the sake of interconnecting subsystems as it will be discussed in detail in Section~\ref{section_composition}. 

In the following, we first introduce the notion of control sub-barrier certificates (CSBC) for dt-SCS with internal inputs and outputs as well as control barrier certificates (CBC) for dt-SCS without internal inputs and outputs. We then utilize those notions to provide some probabilistic upper bounds for solution processes of the large-scale system to reach some unsafe regions over \emph{infinite-time} horizons.

\section{Control (Sub-)Barrier Certificates}
\begin{definition} \label{csbc}
	Consider a dt-SCS $\mathfrak{S}=(X,U,W,\varsigma,f,$ $Y,h)$ with both internal and external inputs and outputs, and sets $X_0, X_u \subseteq X$ as initial and unsafe sets of the system, respectively. A function $\mathds B:X \rightarrow \mathbb{R}_{\geq 0}$ is called a control
	sub-barrier certificate (CSBC) for $\mathfrak{S}$ with respect to initial set $X_0$ and unsafe set $X_u$ if there exists a constant $\eta\in\R_{\geq 0}$ and a symmetric matrix $\underbar X$ with conformal block partitions $\underbar X^{ij}, i,j \in \{1,2\}$ such that
	\begin{align}\label{subsys1}
	&\mathds B(x) \leq \eta,\quad\quad\quad\quad\quad\quad\!\forall x \in X_{0},\\\label{subsys3}
	&\mathds B(x) \geq 1, \quad\quad\quad\quad\quad\quad\!\forall x \in X_{u}, 
	\end{align} 
	and $\forall x\in X$, $\exists \nu\in U$, such that $\forall w\in W$,
	\begin{align} 
	\mathbb{E}&\Big[\mathds B(x(k+1)) \,\,\big|\,\, x(k)\!=\!x, \nu(k)\!=\!\nu, w(k)\!=\!w\Big] \!-\! \mathds B(x(k)) \label{csbceq} \leq \begin{bmatrix}
	w  \\ h(x)\end{bmatrix}^T \begin{bmatrix} \underbar X^{11} & \underbar X^{12} \\ 
	\underbar{X}^{21} & \underbar X^{22} 
	\end{bmatrix}	\begin{bmatrix}
	w  \\ h(x)\end{bmatrix}\!.
	\end{align}
\end{definition}

We now define a similar notion for interconnected dt-SCS without internal inputs and outputs as the following. 

\begin{definition} \label{cbc}
	Consider a dt-SCS $\mathfrak{S}\!=\! (X,U,\varsigma,f)$ without internal inputs and outputs. A function $\mathds B:X \rightarrow \mathbb{R}_{\geq 0}$ is called a control barrier certificate (CBC) for $\mathfrak{S}$ with respect to initial set $X_0$ and unsafe set $X_u$ if there exists constants $\eta \in \mathbb{R}_{\geq 0}$ and $\beta \in \mathbb{R}_{> 0}$ with $\beta > \eta$ such that
	\begin{align}\label{sys2}
	&\mathds B(x) \leq \eta,\quad\quad\quad\quad\quad\!\forall x \in X_{0},\\\label{sys3}
	&\mathds B(x) \geq \beta, \quad\quad\quad\quad\quad\!\forall x \in X_{u},
	\end{align} 
	and $\forall x\in X$, $\exists \nu\in U$, such that
	\begin{align}\label{cbceq}
	\mathbb{E}&\Big[\mathds B(x(k+1)) \,\,\big|\,\, x(k)=x, \nu(k)=\nu\Big]- \mathds B(x(k)) \leq 0.
	\end{align}
	
\end{definition}

\begin{remark}
	The condition $\beta > \eta$ is required in Definition \ref{cbc} for interconnected dt-SCS for the sake of providing meaningful probabilistic guarantees over the satisfaction of safety specifications given by Theorem \ref{probability_bounds}. However, the same condition is not necessarily required in Definition \ref{csbc} for dt-SCS with internal inputs and outputs.
\end{remark}

We now use Definition \ref{cbc} to quantify an upper bound on the probability that the interconnected dt-SCS $\mathfrak{S}$ \textit{reaches} unsafe regions in infinite-time horizon. 

\begin{theorem} \label{probability_bounds}
	Let $\mathfrak{S}=(X,U,\varsigma,f)$ be a dt-SCS without internal inputs and outputs, and $\mathbb{B}$ be a CBC for $\mathfrak{S}$. Then the upper bound on the probability that the solution process of $\mathfrak{S}$ starts from any initial state $a \in X_0$ and reaches an unsafe region $X_u$ under the control policy $\varpi$ 
	is given by
	
	\begin{equation} \label{eq:bounds}
	\mathbb{P}^{a}_{\varpi} \Big\{x(k) \in X_u \text{ for some} \ 0 \leq k < \infty \,\,\big|\,\, a\Big\} \leq \frac{\eta}{\beta}.
	\end{equation}
\end{theorem}

Proof of Theorem \ref{probability_bounds} is provided in the Appendix. The following corollary provides a probability lower bound on the satisfaction of safety specifications, \textit{i.e.}, a probability lower bound with which the solution processes of $\mathfrak{S}$ \textit{avoids} entering unsafe regions in \emph{infinite-time} horizon.

\begin{corollary} \label{lower_bounds}
	Let $\mathfrak{S}=(X,U,\varsigma,f)$ be a dt-SCS without internal inputs and outputs, and $\mathbb{B}$ be a CBC for $\mathfrak{S}$. Then the lower bound on the probability that the solution process of $\mathfrak{S}$ starts from any initial state $a \in X_0$ and avoids entering an unsafe region $X_u$ under the control policy $\varpi$ 
	is given by
	\begin{equation} \label{eq:lower_bounds}
	\mathbb{P}^{a}_{\varpi}\Big\{x(k) \notin X_u \text{ for all } \ 0 \leq k < \infty \,\,\big|\,\, a \Big\} \geq 1 - \frac{\eta}{\beta}.
	\end{equation}
\end{corollary}

\begin{remark} 
	In order to provide probabilistic guarantees in infinite-time horizons, CBC $\mathbb{B}$ in~\eqref{cbceq} is required to be a non-negative supermartingale, \textit{i.e.}, the value of CBC is expected to decay at every time step. This can be quite restrictive and there may not exist a CBC satisfying the supermartingale condition~\eqref{cbceq}. In such a case, it is possible to relax  condition~\eqref{cbceq} by introducing a constant $c > 0$ in the right hand side of~\eqref{cbceq}. In this case, the CBC $\mathbb{B}$ is called $c$-martingale~\cite{steinhardt_finite-time_2012} and condition~\eqref{cbceq} ensures that CBC is decaying with an offset of up to $c$. However, this comes at a cost of providing only finite-time horizon guarantees.
\end{remark}

Unfortunately, finding a CBC for large-scale systems can be difficult due to computational complexity associated with the dimension of the state space. In this article, we consider a large-scale system as an interconnection of several smaller subsystems and develop a compositional scheme to construct the CBC of the interconnected system based on CSBCs of individual subsystems. This is explained in detail in the following section. 

\section{Compositional Construction of CBC for Interconnected Systems} \label{section_composition}

\subsection{Interconnected Stochastic Control Systems}

In this subsection, we provide the formal definition of interconnected discrete-time stochastic control systems as the following. 

\begin{definition}
	Suppose we are given $N \in \mathbb{N}_{\geq 1}$ control subsystems $\mathfrak{S}_i = (X_i,U_i,W_i,\varsigma_i,f_i,Y_{i},h_{i}), i\in \{1,\dots,N\}$, where $X_i \in \mathbb{R}^{n_i}$, $U_i \in \mathbb{R}^{m_i}$, $W_i \in \mathbb{R}^{p_i}$, $Y_{i} \in \mathbb{R}^{r_{i}}$
	along with a matrix $M$ that describes the coupling between the subsystems, with the constraint $M \prod^{N}_{i=1} Y_{i} \subseteq M \prod^{N}_{i=1} W_{i}$ to provide a well-posed interconnection. Then the interconnection of subsystems $\mathfrak{S}_i, i \in \{1,\ldots,N\}$, denoted by $\mathcal{I}(\mathfrak{S}_1,\ldots,\mathfrak{S}_N)$, is the dt-SCS $\mathfrak{S}=(X,U,\varsigma,f)$ such that $X := \prod^{N}_{i=1} X_i$, $U := \prod^{N}_{i=1} U_i$, 
	and $f := \prod^{N}_{i=1} f_i$, with internal inputs constrained according to

	\begin{equation}
	[w_1;\ldots ; w_N]= M[h_{1};\ldots;h_{N}].
	\end{equation}
\end{definition}
\begin{remark}
	For the sake of controller synthesis, we assume that all subsystems $\mathfrak{S}_i, i \in \{1,\ldots,N\}$, have access to their full-state information. The main goal is to synthesize external inputs in order to satisfy specifications over the states of the interconnected system.
\end{remark}
\subsection{Compositional CBC for Interconnected Systems}\label{compositional}

We now provide a compositional framework for obtaining CBC for interconnected dt-SCS  $\mathfrak{S}$ based on CSBCs of subsystems $\mathfrak{S}_i$. Let us assume that there exist a CSBC $\mathds B_i$ as in Definition \ref{csbc} for each control subsystem $\mathfrak{S}_i, i\in \{1,\dots,N\}$, with  $\eta_i \in \mathbb{R}_{\geq 0}$ and symmetric matrix $\underbar X_i$ with conformal block partitions $\underbar X^{11}_i, \underbar X^{12}_i,\underbar X^{21}_i,\underbar X^{22}_i $. We propose the following theorem in order to provide sufficient conditions to obtain a CBC for the interconnected dt-SCS $\mathfrak{S}$ from the CSBCs of subsystems $\mathfrak{S}_i, i \in \{1,\ldots,N\}$. 

\begin{theorem}\label{Thm: Comp}
	Consider an interconnected dt-SCS $\mathfrak{S}=\mathcal{I}(\mathfrak{S}_1,\ldots,$ $\mathfrak{S}_N)$ composed of $N$ control subsystems $\mathfrak{S}_i$, $i \in \{1,\ldots,N\}$, with an interconnection matrix $M$. Assume each control subsystem $\mathfrak{S}_i$ admits a CSBC $\mathbb{B}_i$ with parameter $\eta_i$ according to Definition \ref{csbc}. If
	
	\begin{align}\label{eq:eta}
	\sum_{i=1}^N  \eta_i & < N,\\\label{eq:lmi}
	\begin{bmatrix}
	M \\ I_{\tilde r} 
	\end{bmatrix}^T \underbar X^{comp}& \begin{bmatrix}
	M \\ I_{\tilde r} 
	\end{bmatrix}  \leq 0,
	\end{align} 
	then
	\begin{equation} \label{finalCBC}
	\mathbb{B}(x)=\sum_{i=1}^N  \mathbb{B}_i(x_i)
	\end{equation}
	is a CBC for the interconnected system $\mathfrak{S}=\mathcal{I}(\mathfrak{S}_1,$ $\ldots,\mathfrak{S}_N)$, where 
	\begin{equation} \label{xcomp}
	\underbar X^{comp} := \begin{bmatrix}
	\underbar X_1^{11} &  &  &\underbar X_1^{12} &  &   \\ & \ddots &  &    & \ddots &  & \\
	
	&    &   \underbar X_N^{11}  &    &   &  \underbar X_N^{12} \\
	
	\underbar X_1^{21}  &   &   &  \underbar X_1^{22}  &   &   \\
	& \ddots &  &    & \ddots &  & \\
	&    &   \underbar X_N^{21}  &    &   &  \underbar X_N^{22} \\
	\end{bmatrix}\!,
	\end{equation}
	and $\tilde r= \sum_{i=1}^N r_{i}$ 
	where $r_{i}$ is the dimension of the internal output of subsystem $\mathfrak{S}_i$. 
\end{theorem}
Proof of Theorem \ref{Thm: Comp} is given in the the Appendix.

\begin{remark}\label{compositionality remark}
	Condition~\eqref{eq:lmi} is similar to the linear matrix inequality (LMI) appeared in~\cite{arcak_networks_2016} as a compositional stability condition based on the dissipativity theory. It is shown in~\cite{arcak_networks_2016} that this condition holds independently of the number of subsystems in many physical applications with particular interconnection topologies, e.g., skew symmetric.
\end{remark}

Conventionally, in order to satisfy the compositionality condition in~\eqref{eq:lmi}, the required parameters for sub-barrier certificates (\emph{i.e.,} conditions \eqref{subsys1}-\eqref{csbceq}) are pre-selected and the compositionality condition is checked a posteriori. While this method may provide tractable results for systems with specific interconnection structures, it may not be tractable for networks where structural properties of interconnection topology and subsystems are not apparent. Hence, obtained control sub-barrier certificates may not satisfy the compositionality condition~\eqref{eq:lmi} a posteriori and one needs to redesign them from scratch again. In order to design control sub-barrier certificates while having the compositionality condition~\eqref{eq:lmi} in mind a priori, we employ a distributed optimization method based on an alternating direction method of multipliers (ADMM) algorithm. It allows us to break down a large optimization problem into several smaller sub-problems which can be easier to handle. The solution to the optimization problem provides us with suitable control sub-barrier certificates along with local control policies, satisfying the compositionality condition~\eqref{eq:lmi} and, hence, allowing the computation of control barrier certificates for the interconnected system.

\subsection{Compositional Certification using ADMM Algorithm}

In this subsection, we discuss the ADMM algorithm \cite{boyd_distributed_2011} which allows us to decompose the condition~\eqref{eq:lmi} into local sub-problems and a global one involving the interconnection matrix $M$, as well as matrices $\underbar X_i$, $\forall i\in\{1,\ldots,N\}$, and $\underbar X^{comp}$. Consider an interconnected system $\mathfrak{S}=\mathcal{I}(\mathfrak{S}_1,\ldots,\mathfrak{S}_N)$. The task of constructing CSBC $\mathds B_i$ of subsystems $\mathfrak{S}_i$, $i \in \{1,\ldots,N\}$, can be formulated as a local optimization problem given by 
\begin{align}
\mathcal{S}_i\!=\! \Big\{(\underbar X_i, \eta_i) \, \big | \, \exists~\text{CSBC}~ \mathds B_i ~\text{ w.r.t. conditions}~ \eqref{subsys1}\text{-}\eqref{csbceq}\Big\}. 
\end{align}
Verifying the compositionality condition is a global feasibility problem that can be formulated as
\begin{align}
\mathcal{G}\!=\! \Big\{
(\underbar X_1, \ldots, \underbar X_N,\eta_1,\ldots,\eta_N)
\, \big | \, &\text{conditions} ~\eqref{eq:eta}-\eqref{eq:lmi}~  \text{are satisfied}\Big\}.
\end{align}
We now restate Theorem \ref{Thm: Comp} as a feasibility problem in the following lemma. 
\begin{lemma} \label{lem:admm}
	Consider an interconnected system $\mathfrak{S}=\mathcal{I}(\mathfrak{S}_1,\ldots,\mathfrak{S}_N)$. If there exist matrices $\underbar X_1, \ldots, \underbar X_N$ and constants $\eta_1, \ldots, \eta_N$ such that	
	\begin{align} \notag
	&(\underbar X_i, \eta_i) \in \mathcal{S}_i, \quad \forall i \in \{1,\ldots,N\}, \\ \label{eq:opt}
	&(\underbar X_1, \ldots, \underbar X_N,\eta_1,\ldots,\eta_N) \in \mathcal{G},
	\end{align}
	then $\mathds B(x)=\sum_i^N \mathds B_i(x_i)$ is a CBC for the interconnected system.
\end{lemma}
In order to convert the feasibility problem of Lemma \ref{lem:admm} to an ADMM form, we first define the following indicator functions:
\begin{align*}
\mathbb{I}_{\mathcal{S}_i}(\underbar X_i, \eta_i)= 
&\begin{cases}
0,&(\underbar X_i, \eta_i) \in \mathcal{S}_i,  \\
\infty, & \text{otherwise},
\end{cases} \\
\mathbb{I}_\mathcal{G}(\underbar X_1,\ldots,\underbar X_N, \eta_1 \ldots, \eta_N)=
&\begin{cases}
0, &(\underbar X_1,\ldots, \underbar X_N, \eta_1,\ldots,\eta_N) \in \mathcal{G},  \\
\infty, & \text{otherwise}.
\end{cases}
\end{align*}
Now, by introducing auxiliary variables $\underbar Z_i$ and $\zeta_i$ for each subsystem, we can rewrite \eqref{eq:opt} as an optimization problem in the ADMM form as
\begin{align}\label{eq:admm}
&\text{ADMM}\!:\left\{
\hspace{-0.5mm}\begin{array}{l}\min\limits_{d} ~\!\overset{N}{\underset{i=1}\sum}\mathbb{I}_{\mathcal{S}_i}(\underbar X_i,\eta_i)\!+\!\mathbb{I}_\mathcal{G}(\underbar Z_1,\ldots, \underbar Z_N, \!\zeta_1, \ldots, \!\zeta_N ),\\
\, \text{s.t.} \quad  \,\underbar X_i- \underbar Z_i = 0, \quad \forall i \in \{1,\ldots,N\},\\
\quad\quad\quad\!\eta_i-\zeta_i=0, ~\quad \forall i \in \{1,\ldots,N\},\end{array}\right.
\end{align} 
where $d = (\underbar X_1, \ldots, \underbar X_N,\eta_1, \ldots,  \eta_N,\underbar Z_1, \ldots, \underbar Z_N, \zeta_1, \ldots, \zeta_N)$. In order to decompose large optimization problems into smaller sub-problems, one potential solution, commonly used in the optimization theory, is to split the objective function. Since the first part of the objective function, \emph{i.e.,} $\mathbb{I}_{\mathcal{S}_i}$, is separable by subsystems, one can find a solution parallelly by iterating over $\underbar X_i$, $\eta_i$, $\underbar Z_i$, and $\zeta_i$, alternately with the help of new scaled dual variables $\Lambda_i$ and $\xi_i$ which can take real values over corresponding dimensions. The iterative updating of variables is performed in the following manner:

\begin{itemize}
	\item For each $i \in \{1,\ldots,N\}$, solve the following local problem:
	\begin{align*}
	\underbar X_i^{k+1}, \eta_i^{k+1}= \underset{(\underbar X^*, \eta^*) \in \mathcal{S}_i}{\operatorname{argmin}}\Big\{\|\underbar X^{*}& - \underbar Z_i^k + \Lambda_i^k\|^2_F  + (\eta^*-\zeta_i+\xi_i^k)^2\Big\}.
	\end{align*}
	\item If $\underbar X_{1:N}^{k+1}, \eta_{1:N}^{k+1} \in \mathcal{G}$, then the optimal solution is found and the algorithm can be terminated. If not, we solve the following global problem:
	\begin{align*}
	\underbar Z_{1:N}^{k+1}, \zeta_{1:N}^{k+1}\!=\! \underset{(\underbar Z_{1:N}, \zeta_{1:N}) \in \mathcal{G}}{\operatorname{argmin}}\sum_{i=1}^N\Big\{\|&\underbar X_i^{k+1} - \underbar Z_i + \Lambda_i^k\|^2_F + (\eta_i^{k+1}\!-\!\zeta_i^{*}\!+\!\xi_i^k)^2\Big\}.  
	\end{align*}
	
	\item We update our dual variables as
	\begin{align*}
	&\Lambda_{i}^{k+1}=\underbar X_i^{k+1}- \underbar Z_i^{k+1} + \Lambda_{i}^k, \\ \notag
	&\xi_i^{k+1}=\eta_i^{k+1}-\zeta_i^{k+1} + \xi_i^k,	
	\end{align*}
	and return to the first step until a possible convergence.
\end{itemize}

Optimal solutions of local problems for subsystems $\mathfrak{S}_i$, $i \in \{1,\ldots,N\}$, can be found parallelly by utilizing sum-of-squares optimization (SOS) formulation, as will be discussed later in Subsection \ref{subsec:sos}. This provides us with $\underbar X_i$ that is closest to $\Lambda_i-\underbar Z_i$ in the Frobenius norm and $\eta_i$ that is closest to the value of $\xi_i-\zeta_i$. These values are then passed to the global problem whose solution can be found by using semi-definite programming (SDP). Using the optimal values of both local and global problems, the dual variables $\Lambda_i$ and $\xi_i$ are updated and this procedure is repeated until the solutions converge. Since objective functions of the problem \eqref{eq:admm} are both convex, solutions are guaranteed to converge to optimal ones \cite{meissen_compositional_2015}.

In the next subsection, we describe how to find CSBC of subsystems and compute corresponding local control policies.

\subsection{Computation of CSBC and Control Policy} \label{subsec:sos}

The ADMM algorithm described in the previous subsection requires computation of optimal values of $\underbar X_i$ and $\eta_i$ for subsystems $\mathfrak{S}_i$ such that the  objective function of the local problem is minimized subject to satisfaction of conditions \eqref{subsys1}-\eqref{csbceq}. One can reformulate these conditions as a sum-of-squares (SOS) optimization problem \cite{Parrilo2003} to search for suitable CSBC and corresponding local control policies while computing optimal values of $\underbar X_i$ and $\eta_i$. This can be done by restricting the CSBC to be a non-negative polynomial that can be written as sum-of-squares of different polynomials. To do so, we raise the following assumption.  

\begin{assumption} \label{assumeSOS}
	The stochastic control subsystem $\mathfrak{S}_i$ has a continuous state set $X_i \subseteq \mathbb{R}^{n_i}$, and continuous external and internal input sets $U_i \subseteq \mathbb{R}^{m_i}$ and $W_i \subseteq \mathbb{R}^{p_i}$. Its transition map $f_i: X_{i} \times U_i \times W_i \times \mathcal V_{\varsigma_i}\rightarrow X_i$ is a polynomial function of the state $x_i$, the external input $\nu_i$, and the internal input $w_i$.
\end{assumption} 

Under Assumption \ref{assumeSOS}, conditions~\eqref{subsys1}-\eqref{csbceq} can be reformulated as an SOS optimization problem to search for a polynomial CSBC $\mathds B_i$ and a polynomial controller $\nu_i(\cdot)$ for the subsystem $\mathfrak{S}_i$. The following lemma provides the SOS formulation. 

\begin{lemma}\label{sos}
	Suppose Assumption \ref{assumeSOS} holds and sets $X_i$ , $X_{0_i}$, $X_{u_i}$ can be defined by vectors of polynomial inequalities $X_i=\{x_i \in \mathbb{R}^{n_i} \mid b_i(x_i) \geq 0\}$, $X_{0_i}=\{x_i \in \mathbb{R}^{n_i} \mid b_{0_i}(x_i) \geq 0\}$, and $X_{u_i}=\{x_i \in \mathbb{R}^{n_i} \mid b_{u_i}(x_i) \geq 0\}$, where inequalities are provided element-wise. Similarly, let the internal input set $W_i$ be defined by vectors of a polynomial inequality $W_i=\{w_i\in\R^{p_i}\mid b_{w_i}(w_i)\geq0\}$. Suppose for a given control subsystem $\mathfrak{S}_i$, there exists a sum-of-squares polynomial $\mathds B_i(x_i)$, a constant $\eta_i \in \R_{\geq 0}$,  vectors of sum-of-squares polynomials $\lambda_{0_i}(x_i), \lambda_{u_i}(x_i)$, $\lambda_i(x_i)$, $\lambda_{w_i}(w_i)$ and polynomials $\lambda_{\nu_{ji}}(x_i)$ corresponding to the $j^{\text{th}}$ input in $\nu_i=(\nu_{1_i},\ldots,\nu_{{m_i}_i}) \in U_i \subseteq \mathbb{R}^{m_i}$ of appropriate dimensions such that the following expressions are sum-of-squares polynomials:
	\begin{align} \label{sos1}
	-&\mathds B_i(x_i)-\lambda^{T}_{0_i}(x_i)b_{0_i}(x_i)+\eta_i,\\\label{sos3}
	&\mathds B_i(x_i)-\lambda^{T}_{u_i}(x_i)b_{u_i}(x_i)-1,\\\notag
	-&\mathbb{E}\Big[\mathds B_i(f_i(x_i,\nu_i,w_i,\varsigma_i)) \mid x_i,\nu_i,w_i\Big]\!\!+\!\mathds B_i(x_i)+\begin{bmatrix}
	w \\ h(x)\end{bmatrix}^T \begin{bmatrix} \underbar X^{11} & \underbar X^{12} \\ 
	\underbar{X}^{21} & \underbar X^{22}  
	\end{bmatrix}	\begin{bmatrix}
	w  \\ h(x)\end{bmatrix} \\\label{sos4}
	&+\sum_{j=1}^{m_i}(\nu_{ji}\!-\!\lambda_{\nu_{ji}}(x_i))\!-\!\lambda^{T}_i(x_i)b_i(x_i)\!-\!\lambda_{w_i}^{T}(w_i)b_{w_i}(w_i).
	\end{align}
	Then $\mathds B_i(x_i)$ is a CSBC satisfying conditions~\eqref{subsys1}- \eqref{csbceq} and $\nu_i=[\lambda_{\nu_{1i}}(x_i);\dots;$ $\lambda_{\nu_{m_{ii}}}(x_i)],  \forall i\in\{1,\dots,N\}$, is the corresponding controller for the subsystem $\mathfrak{S}_i$.
\end{lemma}  
Proof of Lemma~\ref{sos} is provided in the Appendix.

\begin{remark}
	Note that one can compute the expected value in~\eqref{sos4} by utilizing the moments of the distribution of $\varsigma_i$ when the distribution of $\varsigma_i$ is known.
\end{remark}

\begin{remark}
	Our proposed computational method is based on SOS optimization in combination with ADMM algorithm and it relies on the assumption that sub-barrier certificates are polynomial. However, there are other methods for the computation of barrier certificates such as the counter-example guided inductive synthesis (CEGIS) \cite{jagtap_formal_2019,jagtap2020compositional} where such an assumption is no longer required at the cost of having more computational complexity.
\end{remark}

\begin{remark}
	The extent of coupling between subsystems can affect the computational complexity in our setting. In particular, if the interconnection topology is too dense (e.g., fully interconnected network), the computational complexity of the ADMM algorithm with LMI and SOS optimization problem potentially increases. 
	More precisely, the complexity of solving LMI \eqref{eq:lmi} is cubic with respect to the number of subsystems $N$. However, under certain sparsity patters in the interconnection topology, one can achieve a linear complexity with respect to $N$ \cite{complexityLMI}. Moreover, the complexity of searching for a CSBC satisfying \eqref{sos4} is polynomial with respect to the number of state and input variables \cite{topcu}. Since dense interconnections create more input variables, they also increase the complexity of searching CSBCs. In general, for a fixed degree of polynomials, one can establish the complexity of the ADMM algorithm to be polynomial with respect to the number of the state and input variables for each subsystem, as well as with respect to the size of the interconnection matrix $M$ (or the number of subsystems $N$) for each iteration.
\end{remark}	
In the next section, we discuss our properties of interest in this work which are $\omega$-regular properties. We then propose a systematic procedure to obtain probabilistic guarantees ensuring the satisfaction of such specifications with the help of barrier certificates. 

\section{Class of Specifications} \label{sec:spec}

The main goal in this work is to synthesize controllers for interconnected dt-SCS ensuring the satisfaction of $\omega$-regular properties \cite{thomas_automata_1991}. Such specifications can be expressed by $\omega$-automata that can recognize infinite words, such as non-deterministic B\"{u}chi automata \cite{buchi_decision_1990}, deterministic Rabin automata \cite{rabin_decidability_1968}, deterministic Streett automata \cite{streett_propositional_1982}, parity automata or Muller automata \cite{muller_infinite_1963}. While the above mentioned automata have different acceptance conditions, they have the same expressive power and all of them recognize $\omega$-regular languages. Here, we use deterministic Streett automata to describe $\omega$-regular properties, whose formal definition is provided as follows. 
\begin{definition} 
	A deterministic Streett automaton (DSA) is a tuple $\mathcal{A}=(Q,$ $q_0, \Sigma, \delta, \emph{Acc})$, where $Q$ is a finite set of states, $q_0 \in Q$ is the initial state, $\Sigma$ is a finite set of input symbols called alphabet, $\delta: Q \times \Sigma \rightarrow Q$ is the transition function and $\emph{Acc}=\{<E_1,F_1>,<E_2,F_2>,\ldots,<E_z,F_z>\}$  refers to the accepting condition of the DSA where $<E_i,F_i>$ with $E_i,F_i \subseteq Q, \forall i \in \{1,\ldots,z\}$, are accepting state pairs. 	
\end{definition}
For the sake of an easier presentation, we define the sets $E=\{E_1,E_2,\ldots,E_z\}$ and $F=\{F_1,F_2,\ldots,F_z\}$ where $<E_i,F_i>\, \in \emph{Acc}$, $\forall i \in \{1,\ldots,z\}$. An infinite sequence of input symbols $\sigma=(\sigma_0, \sigma_1,\ldots) \in \Sigma^{\omega}$ is called an infinite \textit{word} or \textit{trace}. An infinite \textit{run} or \textit{path} $\textbf{q}\!=\!(q_0,q_1,\ldots) \in Q^{\omega}$ on the word $\sigma=(\sigma_0,\sigma_1,\ldots)$ is an infinite sequence of states such that for every $m \geq 0$, we have $q_{m+1}=\delta(q_m,\sigma_m)$. Let $\emph{inf}\,(\textbf{q})$ denote the set of states in $Q$ that are visited infinitely often during the run $\textbf{q}$. Then $\textbf{q}$ is said to be an accepting run if for all $E_i \in E$ and $F_i \in F$, $i \in \{1,\ldots,z\}$, we have $\emph{inf}\,(\textbf{q}) \wedge E_i = \emptyset$ or $\emph{inf}\,(\textbf{q}) \wedge F_i \neq \emptyset$, and the corresponding word $\sigma(\textbf{q})$ is said to be accepted by the DSA $\mathcal{A}$, denoted by $\sigma(\textbf{q}) \models \mathcal{A}$.  
The language of $\mathcal{A}$, denoted by $\mathcal{L}(\mathcal{A})$, comprises all the words accepted by $\mathcal{A}$. 

We consider specifications expressed by accepting languages of DSA $\mathcal{A}$ when input symbols are defined over a set of atomic propositions $\mathcal{AP}$ as the alphabet, \textit{i.e.}, $\Sigma=2^{\mathcal{AP}}$. For instance, specifications expressed as linear temporal logic (LTL) formulae can be represented by DSA with the help of existing tools like \texttt{ltl2dstar} \cite{klein_ltl2dstar}. 

\begin{remark}
	A DSA $\mathcal{A}$ with the set $E=\emptyset$ accepts any infinite run. Therefore, without loss of generality, we assume that the set $E$ is non-empty.
\end{remark}

It should be noted that while deterministic B\"{u}chi automata are also a class of $\omega$-automata and are used for representing languages over infinite words, their expressive power is strictly weaker than other classes of $\omega$-automata such as non-deterministic B\"{u}chi, deterministic Streett or Rabin automata. It is worth mentioning that if the verification is the main objective, one can utilize both deterministic and non-deterministic B\"{u}chi automata. The latter is preferred because it has higher expressive power, and accordingly, it can represent $\omega$-regular properties. However, in our work, we deal with controller synthesis problem where the determinism of automata is crucial, and one cannot directly work with non-deterministic B\"uchi automata (NBA). In this case, one needs to determinize the automata without losing their expressiveness. Although the idea of this paper can be applied to less expressive automata including deterministic B\"uchi automata, we prefer to deal with the full class of LTL properties. Hence, we work with deterministic Streett automata since they have the same expressive power as NBA.

\subsection{Satisfaction of Specifications by Interconnected Systems}

Here, we define how $\omega$-regular properties are connected to solution processes of interconnected dt-SCS $\mathfrak{S}=(X,U,\varsigma,f)$ via a measurable labeling function $L: X \rightarrow \mathcal{AP}$.

\begin{definition} \label{def:label}
	For an interconnected dt-SCS $\mathfrak{S}=(X,U,\varsigma,f)$ and a DSA $\mathcal{A}=(Q,q_0,\mathcal{AP},\delta,\emph{Acc})$, consider a labeling function $L: X \rightarrow \mathcal{AP}$. For an infinite-state sequence $\textbf{x}=(x(0),x(1),\ldots) \in X^\omega$, the corresponding word over $\mathcal{AP}$ is given by $L(x) := (\sigma_0, \sigma_1,\ldots) \in \mathcal{AP}^{\omega}$\!, where $\sigma_i=L(x(i))$ for all $i \in \mathbb{N}$.
\end{definition}

The set of atomic propositions $\mathcal{AP}=\{p_0,p_1,\ldots, p_R\}, R \in \mathbb{N}$, provides a measurable partition of the state space $X=\bigcup_{i=1}^{R} X^{i}$ via the labeling function $L:X \rightarrow \mathcal{AP}$ such that $L(x \in X^{i})=p_i$. Without loss of generality, it can be assumed that $X^{i} \neq \emptyset$ for any $i \in \{1,\ldots,R\}$. 

\begin{remark}
Since $\mathcal{AP}$ provides a measurable partition of the state set $X$, for any two atomic propositions $p_i, p_j \in \mathcal{AP}$, $i \neq j$, $i,j \in \{1,\ldots,R\}$, we have that $p_i \wedge p_j = \emptyset$.	Therefore, while constructing the DSA $\mathcal{A}$ corresponding to the required specification, one can remove the edges with $p_i \wedge p_j$ as they are infeasible. Moreover, other Boolean combinations of atomic propositions may also be resolved. For example, an edge with $p_i \vee p_j$ can be resolved by adding two new edges with $p_i$ and $p_j$, respectively. The negation  $\neg p_i$ can also be handled in a similar fashion. Therefore, we assume in the remainder of the paper that the alphabet $\Sigma$ is defined directly over the set of atomic propositions rather than its power set, \textit{i.e.}, $\Sigma=\mathcal{AP}$.
\end{remark}

We now define the probability with which solution processes of interconnected dt-SCS $\mathfrak{S}$ defined in \eqref{Eq_2a} satisfy an $\omega$-regular specification represented by DSA $\mathcal{A}$.

\begin{definition}
	Consider an interconnected dt-SCS $\mathfrak{S}=(X,U,\varsigma,f)$, a specification given by the accepting language of the DSA $\mathcal{A}=(Q,q_0,\mathcal{AP},\delta,\emph{Acc})$ and a labeling function $L:X \rightarrow \mathcal{AP}$. Then the probability with which the solution process $x^{a\varpi}$ under the control policy $\varpi$ with an initial condition $x(0)=a$ satisfies the specification expressed by $\mathcal{A}$ is given by $\mathbb{P}\{L(\textbf{x}^{a\varpi}) \models \mathcal{A} \}$.
\end{definition}

In order to tackle the synthesis problem considered in this paper, we should compute a control policy along with a tight lower bound on the probability that the interconnected dt-SCS satisfies a specification expressed by DSA. This is formally stated as follows.

\begin{resp}
	\begin{problem} \label{probstate}
		Given a dt-SCS $\mathfrak{S}=(X,U,\varsigma,f)$, a specification represented by the accepting language of a DSA  $\mathcal{A}=(Q,q_0,\mathcal{AP},\delta,\textit{Acc})$ over a set of atomic propositions $\mathcal{AP}=\{p_0,p_1\ldots,p_R\}$,  $R \in \mathbb{N}$, and a labeling function $L:X \rightarrow \mathcal{AP}$, compute a control policy $\varpi$ (if existing) and a constant $\epsilon \in [0,1]$ such that $\mathbb{P}\{L(\textbf{x}^{a\varpi}) \models \mathcal{A} \} \geq \epsilon$.
	\end{problem}
\end{resp}

Two underlying challenges make this problem difficult to tackle. First and foremost, since the dimension of the system is potentially very high, computing a CBC for a large-scale system is computationally intractable. To tackle this issue, we consider a dt-SCS $\mathfrak{S}$ (without internal inputs and outputs) as an interconnection of smaller subsystems $\mathfrak{S_i}$ (with internal inputs and outputs), and utilize the proposed compositional framework to construct a CBC of the interconnected system based on CSBCs of subsystems, as discussed in Section~\ref{section_composition}. The second difficulty is due to the complex specification. The notion of CBCs as given in Definition \ref{cbc} only allows us to provide guarantees over safety specifications. We now extend this notion to cover properties represented by DSA. To do this, we decompose the DSA into a set of sequential safety specifications, such that the satisfaction of these safety specifications lead to the satisfaction of the original DSA. Then, our problem reduces to computing the corresponding CBCs and appropriate control policies for these safety specifications. 

Let sets $X_0$ and $X_u$, as introduced in Definition~\ref{cbc}, be connected to the atomic proposition $\mathcal{AP}$ via some labeling function $L: X \rightarrow \mathcal{AP}$. We assume that these sets are decomposed as $X_0= \prod_{i=1}^{N} X_{0_i}$ and $X_u=\prod_{i=1}^{N} X_{u_i}$. In other words, sets $X_0$ and $X_u$ can be written as Cartesian products of their counterparts for subsystems. This implies that sets $X_{0_i}$ and $X_{u_i}$, $i \in \{1,\ldots,N\}$, are connected to the corresponding decomposed structure of the atomic proposition $\mathcal{AP}$ via the same labeling function. Now, given an interconnected dt-SCS $\mathfrak{S}$ and a desired specification as a DSA $\mathcal{A}$, our aim is to first decompose negation of the specification into simple reachability problems. We then compute CBC along with a suitable control policy compositionally via CSBC and local control policies of subsystems by utilizing Theorem \ref{Thm: Comp}. Accordingly, we obtain probabilistic upper bounds for these reachability tasks, which can then be combined to acquire an overall lower bound on the probability that solution processes of the interconnected system satisfy the original specification.
Next, we describe the sequential reachability decomposition method for specifications expressed by DSA, inspired by \cite{topcu}.

\subsection{Specification Decomposition}
\label{subsec:seq_reach}
In order to facilitate controller synthesis for general $\omega$-regular specifications represented by accepting languages of DSA, we use a divide-and-conquer method and reduce the automaton to a set of simple sequential safety tasks.
Consider a DSA $\mathcal{A}=(Q,q_0,\mathcal{AP},\delta,Acc)$ which expresses the desired specification for dt-SCS $\mathfrak{S}$. Note that in order for the dt-SCS $\mathfrak{S}$ to satisfy the specification expressed by $\mathcal{A}$, the words $L(\textbf{x}^{a\varpi})$ corresponding to the solution processes $\textbf{x}^{a\varpi}$ of $\mathfrak{S}$ must be accepted by $\mathcal{A}$. This means that the corresponding runs of the form $\textbf{q}=(q_0,q_1,\ldots) \in Q^{\omega}$ must satisfy the following condition: for all $E_i \in E$ and $F_i \in F$, $i \in \{1,\ldots,z\}$, $\emph{inf}\,(\textbf{q}) \wedge E_i = \emptyset$ or $\emph{inf}\,(\textbf{q}) \wedge F_i \neq \emptyset$. Note that satisfying $\emph{inf}\,(\textbf{q}) \wedge E_i = \emptyset$, for all $E_i \in E$, automatically implies the satisfaction of the original acceptance condition of the DSA $\mathcal{A}$. We refer to this as the partial acceptance condition of the DSA. Moreover, we call an infinite run $\bar{\textbf{q}}=(q_0,q_1,\ldots)$ a partially accepting state run iff for all $E_i \in E$, we have $\emph{inf}\,(\bar{\textbf{q}}) \wedge E_i = \emptyset$. 
	
Now, we provide the decomposition of the DSA $\mathcal{A}$ into consecutive safety tasks, such that the satisfaction of these safety tasks leads to the satisfaction of the partial acceptance condition of $\mathcal{A}$. To do so, we first obtain all the \textit{partially accepting lasso runs} (or simply, \textit{lassos}) of $\mathcal{A}$. Such a lasso consists of a simple (\textit{i.e.} without self-loops) finite path from the initial state $q_0 \in Q$ to a state in $E$, concatenated with a simple finite cycle from the state in $E$ to itself. Then, formally, a lasso is a pair $\tilde{\textbf{q}}=(\tilde{\textbf{q}}_f,\tilde{\textbf{q}}_l)$ such that $\tilde{\textbf{q}}_f=(q_0^f,q_1^f,\ldots,q_{a_f}^f,q_{0}^l)$ represents the finite path and $\tilde{\textbf{q}}_l=(q_0^l,q_1^l,\ldots,q_{a_l}^l,q_0^l)$ represents the finite cycle, where $a_f, a_l \in \mathbb{N}$, $q_0^f = q_0$ and $q_0^l \in E$. Note that the number of such lassos for the DSA $\mathcal{A}$ is finite since $\mathcal{A}$ consists of finite numbers of states and edges. Now, we define a set $\mathcal{R}$ to be the set of all such lassos, \textit{i.e.},
\begin{align}\notag
\mathcal{R} := \big\{\tilde{\textbf{q}}= & (q_0^f,q_1^f,\ldots,q_{a_f}^f,q_0^l,q_1^l,\ldots,q_{a_l}^l,q_0^l)\,\big |\,  q_0^f=q_0, q_0^l \in E, q_r^f \neq q_{r+1}^f, r \leq a_f,~ \\ \notag &\hspace{15em} \text{and }  q_{s}^l \neq q_{s+1}^l, \forall s \leq a_l\big\}.
\end{align}
Moreover, we define the set $\mathcal{R}_f$ containing only the simple finite paths as $\mathcal{R}_f := \{ \tilde{\textbf{q}}_f=(q_0^f,q_1^f,\ldots,q_{a_f}^f,q_{0}^l) \mid q_0^f=q_0, q_0^l \in E$ $\text{ and } q_r^f \neq q_{r+1}^f, r \leq a_f\}$, and the set $\mathcal{R}_l$ to contain only the simple finite cycles as $\mathcal{R}_l := 
\{\tilde{\textbf{q}}_l=(q_0^l,q_1^l,\ldots,q_{a_l}^l,q_0^l) \mid q_0^l \in E, \text{and } q_{s}^l \neq q_{s+1}^l, \forall s \leq a_l\}$.
Now, for each $p \in \mathcal{AP}$, we define a set $\mathcal{R}^p$ as 
\begin{align}
\mathcal{R}^p :=\big\{\tilde{\textbf{q}}=  &(q_0^f,q_1^f,\ldots,q_{a_f}^f,q_0^l,q_1^l,\ldots,q_{a_l}^l,q_0^l)\in \mathcal{R} \,\big |\, \sigma(q_0^f,q_1^f) = p \in \mathcal{AP}\big\}.
\end{align}
Similarly, sets $\mathcal{R}^p_f$ and $\mathcal{R}^p_l$ are defined for simple finite paths and simple finite cycles, respectively. Now, in order to perform decomposition into safety task, we define a set $\mathcal{P}^p(\tilde{\textbf{q}})$ for any $\tilde{\textbf{q}}=(q_0,q_1,\ldots, q_{a_f+a_l+3})$ as  
\begin{equation} \label{eq:reach}
\mathcal{P}^p(\tilde{\textbf{q}})=\big\{(q_i,q_{i+1},q_{i+2}) \,\big |\, 0 \leq i \leq a_f+a_l+1\big\}.
\end{equation} 
Each element in $\mathcal{P}^p(\tilde{\textbf{q}})$ has a length of $3$ and corresponds to a safety task. Consequently, we define $\mathcal{P}^p(\tilde{\textbf{q}}_f)$ and $\mathcal{P}^p(\tilde{\textbf{q}}_l)$ to comprise the safety tasks obtained from simple finite paths and cycles $\tilde{\textbf q}_f$ and $\tilde{\textbf q}_l$, respectively, for each $p \in \mathcal{AP}$. Finally, we define $\mathcal{P}(\mathcal{A})= \bigcup_{p \in \mathcal{AP}} \bigcup_{\tilde{\textbf{q}} \in \mathcal{R}^p}\mathcal{P}^p(\tilde{\textbf{q}})$ as the set of all such safety tasks arising from finite fragments in DRA $\mathcal{A}^c$.

\begin{remark} \label{self-loops}
	Note that even though self-loops are ignored while decomposing the DSA $\mathcal{A}$ into safety tasks $\vartheta=(q_i,q_{i+1},q_{i+2}) \in \mathcal{P}^p(\tilde{\textbf{q}})$, it is crucial to account for the time spent in the self-loops before reaching the state $q_{i+2}$ from $q_i$. This is automatically accounted for via the construction of control barrier certificates as in Definition~\ref{cbc} (cf. Lemma~\ref{lem:cbc}).  		

\end{remark}

We employ the following example for the sake of better illustration. 

\begin{figure} 
	\begin{center}
		\includegraphics[scale=1]{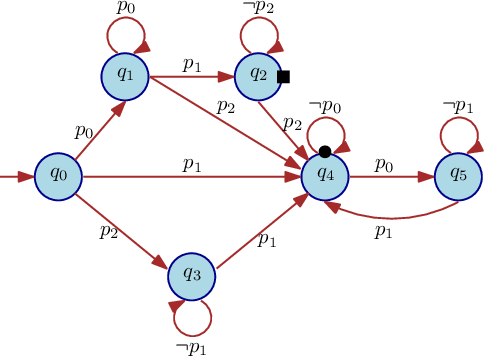} 
		\caption{DSA $\mathcal{A}$ employed in Example~\ref{example}.}
		\label{fig:DRAexample}
	\end{center}
\end{figure}

\begin{example} \label{example}
	We perform safety decomposition for the DSA $\mathcal{A}$ shown in Figure \ref{fig:DRAexample}. The figure indicates with an arrow $\rightarrow$ the initial state of the system, while $\newmoon$ and $\blacksquare$ indicate the states that can be visited finitely and infinitely many times, respectively. In other words, we have $q_0$ as initial state, the set of the atomic proposition $\mathcal{AP}=\{p_0,p_1,p_2\}$ and $\emph{Acc}=<q_4,q_2>$ as the acceptance condition. Therefore, an infinite run $\textbf{q}$ is accepted if it visits $q_4$ only finitely often or $q_2$  infinitely often. In order to decompose the problem into safety tasks, we consider the partially accepting lasso runs of the DSA $\mathcal{A}$ and obtain the set $\mathcal{R}$ consisting of all such lassos. This is given by
	\begin{align*}
	\mathcal{R}=\big\{&(q_0,q_4,q_5,q_4),(q_0,q_3,q_4,q_5,q_4),(q_0,q_1,q_4,q_5,q_4), \\ &(q_0,q_1,q_2,q_4,q_5,q_4)\big\}.
	\end{align*} 
	The sets $\mathcal{R}^p$ for $p \in \mathcal{AP}$, are obtained as 
	\begin{align*}
	&\mathcal{R}^{p_0}=\big\{(q_0,q_1,q_4,q_5,q_4),(q_0,q_1,q_2,q_4,q_5,q_4)\big\}, \\ &\mathcal{R}^{p_1}=\big\{(q_0,q_4,q_5,q_4)\big\}, \quad
	\mathcal{R}^{p_2}=\big\{(q_0,q_3,q_4,q_5,q_4)\big\}.
	\end{align*}
	Now for each $\tilde{\textbf{q}} \in \mathcal{R}^p$, we define $\mathcal{P}^p(\tilde{\textbf{q}})$ as follows:
	\begin{align*}
	&\mathcal{P}^{p_0}(q_0,q_1,q_2,q_4,q_5,q_4)=\big\{(q_0,q_1,q_2),(q_1,q_2,q_4),(q_2,q_4,q_5),(q_4,q_5,q_4)\big\}, \\
	&\mathcal{P}^{p_0}(q_0,q_1,q_4,q_5,q_4)=\big\{(q_0,q_1,q_4),(q_1,q_4,q_5),(q_4,q_5,q_4)\big\}, \\
	& \mathcal{P}^{p_1}(q_0,q_4,q_5,q_4)=\big\{(q_0,q_4,q_5),(q_4,q_5,q_4)\big\}, \\  &\mathcal{P}^{p_2}(q_0,q_3,q_4,q_5,q_4)=\big\{(q_0,q_3,q_4),(q_3,q_4,q_5),(q_4,q_5,q_4)\big\}.
	\end{align*}	
	Finite words $\sigma(\tilde{\textbf{q}})$ corresponding to $\tilde{\textbf{q}} \in \mathcal{R}^p$ are obtained as
	\begin{align*}
	&\sigma(q_0,q_1,q_2,q_4,q_5,q_4)=(p_0,p_1,p_2,p_0,p_1), \\ &\sigma(q_0,q_1,q_4,q_5,q_4)=(p_0,p_2,p_0,p_1), \quad  \sigma(q_0,q_4,q_5,q_4)=(p_1,p_0,p_1),\\ 
	&\sigma(q_0,q_3,q_4,q_5,q_4)=(p_2,p_1,p_0,p_1).
	\end{align*}
\end{example}
We now propose a systematic procedure utilizing CBC to obtain a suitable control policy while computing (preferably maximizing) the lower bound on the probability that interconnected dt-SCS $\mathfrak{S}$ satisfies the specification expressed by the DSA $\mathcal{A}$. To do this, we first consider all the elements in the set $\mathcal{P}(\mathcal{A})$, each of which characterizes a safety task. We the compute the upper bound on the probability that these safety tasks are violated, and then combine them to obtain an overall lower bound on the probability of the satisfaction of the specification given by $\mathcal{A}$.

As a direct consequence of Theorem \ref{probability_bounds}, we propose the following lemma to obtain CBC and upper bounds on probabilities of violating these safety tasks. 

\begin{lemma} \label{lem:cbc}
	For a lasso $\tilde{\textbf{q}} \in \mathcal{R}^p$ with $p \in \mathcal{AP}$, consider a safety task $\vartheta=(q,q',q'') \in \mathcal{P}^p(\tilde{\textbf{q}})$. If there exists a CBC and a suitable control policy $\varpi$ such that conditions \eqref{sys2}-\eqref{cbceq} are satisfied with $X_0=L^{-1}(\sigma(q,q'))$, $X_u=L^{-1}(\sigma(q',q''))$ and constants $\eta_\vartheta \in \mathbb{R}_{\geq 0}$ and $\beta_\vartheta \in \mathbb{R}_{>0}$, then the probability that the solution process of dt-SCS $\mathfrak{S}$ with initial condition $a \in X_0$ reaches the region $X_u$ under the policy $\varpi$ is upper bounded with $\frac{\eta_\vartheta}{\beta_\vartheta}$ as obtained in~\eqref{eq:bounds}. 

\end{lemma}

\begin{remark} \label{rem:ignoreF}
	The satisfaction of the specification represented by the DSA $\mathcal{A}$ requires the disjunction of two different occurences, \textit{i.e.}, states in $E$ should be visited finitely often or states in $F$ should be visited  infinitely often. However, since the disjunction is already satisfied when one of the occurences holds, the probability of satisfaction of the specification represented by $\mathcal{A}$ can be ultimately lower bounded by the probability of the states in $E$ being visited only finitely often. Therefore, we can ignore states in $F$ and proceed with sequential decomposition only by taking into account states in $E$. This is tailored to the nature of CBCs which provide safety guarantees and results in some conservatism in our approach (cf. Section~\ref{subsec:disc}).
\end{remark}

\begin{remark} \label{rem:noCBC}
	For any $\vartheta=(q,q',q'')$, if we have $L^{-1}(\sigma(q,q')) \cap L^{-1}(\sigma(q',q'')) \neq \emptyset$, then the safety task does not admit any control barrier certificate, and correspondingly the probability of violating the safety task is $1$. This is due to the conflict between conditions \eqref{sys2} and \eqref{sys3}.
\end{remark}

\section{Control Policy and Probability Computation}

Generally, every safety task of $\mathcal{P}(\mathcal{A})$ admits a single CBC and its corresponding control policy. However, in a scenario where there is more than one edge emanating from a single state in the automaton, this can result in ambiguities. For this reason, we combine multiple safety tasks into a single partition set and adopt a switching control policy dependent on the location in the automaton. The next subsection explains the switching control policy in detail. We also discuss the computation of the overall lower bound on the probability that solution processes of the interconnected system satisfy the original specification. 

\subsection{Control Policy} \label{control_policy}

Consider the DSA $\mathcal{A}$ shown in Figure \ref{fig:DRAexample}. Consider two safety tasks from the set $\mathcal{P}(\mathcal{A}^c)$ as $\vartheta_1=(q_0,q_1,q_2)$ and $\vartheta_2=(q_0,q_1,q_4)$. Ideally, we must compute two different CBCs and control policies for each of these tasks, one for avoiding the region $L^{-1}(p_1)$ and the other for avoiding $L^{-1}(p_2)$ from a \emph{common} initial region $L^{-1}(p_0)$. Since one cannot employ two different controllers simultaneously in the same region of the state space, this issue results in ambiguity while deploying controllers for the closed-loop system. We resolve this issue by combining the two safety tasks into one by simply replacing the set $X_u$ in Lemma \ref{lem:cbc} with the union of regions corresponding to the alphabet present in all outgoing edges from the common state. To do so, we combine all safety tasks in $\mathcal{P}(\mathcal{A})$ with a common CBC and put them together in a single partition set. Such sets are defined as 
\begin{align*}
\gamma_{(q,q',\Delta(q'))} \!:=\! \big\{(q,q',q'') \in \mathcal{P}(\mathcal{A})\,\big|\, q,q',q'' \in Q\text{ and }  q'' \in \Delta(q')\big\},
\end{align*}
where $\Delta(q)$ is the set of states that can be reached from a state $q \in Q$. For the partition set $\gamma_{(q,q',\Delta(q'))}$, the corresponding CBC and control policy are denoted as $\mathds{B}_{\gamma_{(q,q',\Delta(q'))}}(x)$ and $\nu_{\gamma_{(q,q',\Delta(q'))}}(x)$, respectively. For all safety tasks $\vartheta \in \mathcal{P}({\mathcal{A}})$, we therefore have
\begin{align} \nonumber
\mathds{B}_\vartheta(x)=\mathds{B}_{\gamma_{(q,q',\Delta(q'))}}(x) \text{ and } \nu_\vartheta(x)=\nu_{\gamma_{(q,q',\Delta(q'))}}(x),  
\ \text{ if } \vartheta \in \gamma_{(q,q',\Delta(q'))}. \nonumber
\end{align}
The system admitting a switching control policy as the control input depending on the state of the automaton. To represent such a switching control policy, a new switching automaton $\mathcal{A}_s$ is constructed. This method has been adapted from \cite{jagtap_formal_2019} where switching policy was obtained in the context of DFA. 

For the DSA $\mathcal{A}=(Q,q_0,\mathcal{AP},\delta,\emph{Acc})$, we represent the corresponding switching mechanism as $\mathcal{A}_s=(Q_{s},q_{0s},\mathcal{AP}_s,$ $\delta_s)$ where
$Q_s := q_{0s} \cup \{(q,q',\Delta(q')) \ | \ q,q' \in Q\}$ is the set of states, $q_{0s}:= (q_0,\Delta(q_0))$ is the initial state and $\mathcal{AP}_{s}=\mathcal{AP}$ is the set of atomic proposition. The transition function $\delta_{s}$ is defined as
\begin{itemize}
	\item for $q_{0s}=(q_0,\Delta(q_0))$, we have $\delta_s((q_0,\Delta(q_0)),\sigma_{(q_0,q_0')})=(q_0,q_0',\Delta(q_0'))$ such that $q_0' \in \Delta(q_0)$;
	\item for all $q_s=(q,q',\Delta(q')) \in Q_s \backslash q_{0s}$, we have $\delta_s((q,q',\Delta(q'), \sigma_{(q',q'')})=(q',q'',\Delta(q''))$ such that $q,q',q'' \in Q$, $q'' \in \Delta(q')$.
	
\end{itemize}

Finally, one can obtain the control policy for Problem \ref{probstate} as
\begin{equation} \label{eq:controlpolicy}
\varpi(x,q_s)=\nu_{\gamma_{q'_s}}(x), \ \  \forall(q_s,L(x),q'_s) \in \delta_s.
\end{equation}

\begin{figure} 
	\begin{center}
		\includegraphics[scale=0.7]{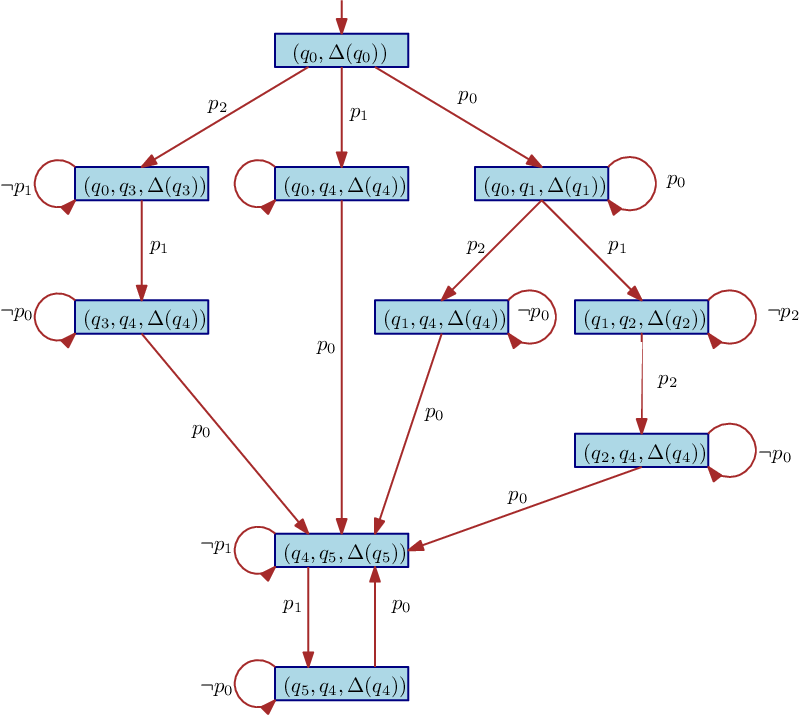} 
		\caption{Automaton $\mathcal{A}_s$ representing switching mechanism.}
		\label{fig:switchingdra}
	\end{center}
\end{figure}

{\bf Example~\ref{example} (continued.)} The automaton $\mathcal{A}_s$ representing the switching control policy for Example \ref{example} is shown in Figure \ref{fig:switchingdra}.

\subsection{Probability Computation}

We now compute the lower bound on the probability that the interconnected dt-SCS $\mathfrak{S}$ satisfies the desired specification expressed by the DSA $\mathcal{A}$. This is done by first computing the upper bounds on the probability of violating the safety tasks $\vartheta=(q,q',q'') \in\mathcal{P}(\mathcal{A}$) using Lemma~\ref{lem:cbc} and combining them to obtain the probability upper bound on visiting the states in $E$ infinitely often. This is then used to compute the probability lower bound on visiting the states in $E$ finitely often, thereby providing the lower bound on the probability with which the interconnected dt-SCS $\mathfrak{S}$ satisfies the specification. This is formally explained in the following theorem.  
\begin{theorem} \label{thm:sumprod}
For a specification expressed by a DSA $\mathcal{A}$, let $\mathcal{R}^p, \mathcal{R}^p_f$, and $\mathcal{R}^p_l$ be all lassos, simple finite paths and simple finite cycles for $p \in \mathcal{AP}$, respectively. Moreover, let $\mathcal{P}^p(\tilde{\textbf{q}}),\mathcal{P}^p(\tilde{\textbf{q}}_f),$ and $\mathcal{P}^p(\tilde{\textbf{q}}_l)$ be the set of safety tasks derived from $\mathcal{R}^p, \mathcal{R}^p_f$, and $\mathcal{R}^p_l$, respectively. The lower bound on the probability that the solution processes of the dt-SCS  $\mathfrak{S}$ start from an initial state $a \in L^{-1}(p)$ and satisfy the specification represented by $\mathcal{A}$ is given by
\begin{equation} \label{eq:sumprod}
\mathbb{P}\big\{L(x^{a\varpi}) \models \mathcal{A}\big\} \geq 1 - \!\! \sum_{\tilde{\textbf{q}} \in \mathcal{R}^p} 	
\prod_{\vartheta \in \mathcal{P}^p(\tilde{\textbf{q}})} \! \! 
\begin{cases} 
\frac{\eta_\vartheta}{\beta_\vartheta} \hspace{-0.5em} &  \vartheta\!\notin\! \mathcal{P}^p(\tilde{\textbf{q}}_l), \\
0  & \vartheta\!\in\! \mathcal{P}^p(\tilde{\textbf{q}}_l) \text{ and }  \eta_\vartheta < \beta_\vartheta, \\
1 &  \vartheta\!\in\! \mathcal{P}^p(\tilde{\textbf{q}}_l) \text{ and } \eta_\vartheta \nless \beta_\vartheta.
\end{cases}
\end{equation}
\end{theorem}
Proof of Theorem \ref{thm:sumprod} is provided in the Appendix.
\begin{remark} \label{rem:loopprob}
	Note that if any safety task $\vartheta=(q,q',q'') \in \mathcal{P}^p(\tilde{\textbf{q}}_l)$ admits a CBC, then $\eta_\vartheta < \beta_\vartheta$  and correspondingly the probability upper bound of reaching a state in $E$ \b{infinitely} many times following the loop $\tilde{\textbf{q}}_l$ is $0$. This is because the existence of CBC guarantees that probability of the loop $\tilde{\textbf{q}}_l$ being taken is less than $1$, and correspondingly, the probability of those loops being taken infinitely often becomes $0$. However, if no CBC exists for such $\vartheta \in \mathcal{P}^p(\tilde{\textbf{q}}_l)$, then one has to consider CBCs for $\vartheta \notin \mathcal{P}^p(\tilde{\textbf{q}}_l)$. In such a case, the probability of visiting states in $E$ only finitely often is lower bounded by the probability of visiting those states at most once.
\end{remark}
\begin{remark} \label{rem:Ntimes}
	Note that we only provide probabilistic guarantees for visiting the states in $E$ at most once in the case that the safety tasks in finite cycles do not admit CBCs. This leads to some conservatism in our approach. However, one can also obtain probabilistic guarantees for visiting the states in $E$ at most $N \in \N$ times. To do this, a DSA $\mathcal{A}$ is reconstructed by duplicating the states in $E$ and the states reachable from the ones in $E$ for $N$ number of times and adding extra transitions to these states such that the language of the reconstructed DSA remains the same. This is illustrated for Example~\ref{example} in Figure 3 for $N=2$, where additional states $q'_4$ and $q'_5$ are added by duplicating the states $q_4$ and $q_5$ respectively, such that $q'_4 \in E$. Then, by ensuring that $q'_4$ is visited at most once in the reconstructed DSA, we accordingly provide guarantees for the state $q_4$ in the original DSA $\mathcal{A}$ (Figure~\ref{fig:DRAexample}) to be visited at most twice. Note that the formal definition of such reconstruction is omitted for the sake of simple presentation. 
\end{remark}

\begin{figure}
	\begin{center}
	\includegraphics[scale=0.9]{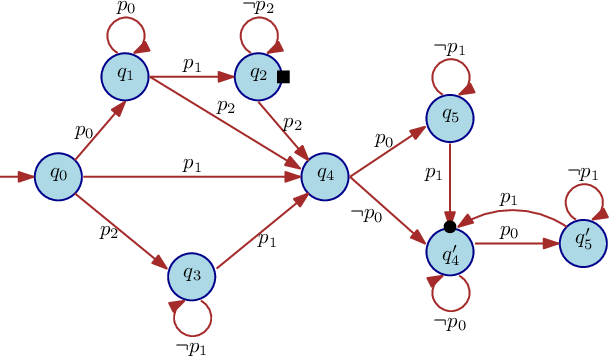}
	\caption{Reconstruction of DSA $\mathcal{A}$ according to Remark~\ref{rem:Ntimes}}
	\end{center}
		\label{fig:ntimes}
\end{figure}

\begin{remark}
	Note that if a safety task given by $\vartheta \in \mathcal{P}(\mathcal{A})$ does not admit a CBC, the probability lower bound for that safety task is considered to be $0$. To obtain potentially a \emph{non-trivial} probability lower bound for the satisfaction of the original property, at least one safety task should have a suitable CBC.
\end{remark}

\begin{remark}
	The probabilistic lower bound obtained in Theorem \ref{thm:sumprod} may decrease to $0$ if the probability upper bounds of violating individual safety tasks obtained from Lemma~\ref{lem:cbc} are either very high (close to $1$) or trivial (equal to $1$). In addition, the number of subsystems can potentially affect the probabilistic guarantee depending on the values of $\eta_i$ obtained in \eqref{sos1} for each subsystem. For instance, when subsystems are identical (i.e., $\eta_i$ obtained for subsystems are equal), the upper bound in \eqref{eq:bounds} remains unchanged w.r.t. the number of subsystems and is always equal to $\eta_i$ since $\eta=N\eta_i$ and $\beta=N$ (see case study). When subsystems are not identical, the upper bound in \eqref{eq:bounds} is always equal to the average of $\eta_i$, i.e., $\frac{\sum_i^N \eta_i}{N}$. In this case, adding more subsystems can either increase or decrease the lower bound.	
\end{remark}	

\begin{remark}
	Trivial probability of $0$ and arbitrary control policy is possible only in the worst-case scenario where our algorithm fails to compute CBCs for all safety tasks in the DSA. Note that safety tasks in the finite cycles of the lassos play a crucial role in obtaining tight lower bounds, as mentioned in Remark~\ref{rem:loopprob}. Therefore, it is beneficial to first search for suitable CBCs and corresponding control policies for safety tasks $\vartheta \in \mathcal{P}^p(\tilde{\textbf{q}}_l)$ in order to obtain tight lower bounds. Moreover, the ADMM algorithm proposed here aims at minimizing the value of $\eta_i$ for subsystems as per equation \eqref{eq:admm}, thereby minimizing overall $\eta$ in equation \eqref{eq:bounds} and allowing the maximization of $\epsilon$ according to Theorem~\ref{thm:sumprod}. This ensures that the probability bounds are (potentially) as tight as possible. 
\end{remark}

\begin{remark}
	While the computation of probabilities of satisfaction for DSA specifications via Theorem~\ref{thm:sumprod} requires the enumeration of all the lassos, the maximum number of CBCs and associated probabilities we need to compute depends on the cardinality of the set of atomic propositions. For instance, if the set of atomic propositions has only three elements, we only require computing a maximum of six CBCs independently of the structure of DSA.
\end{remark}	

\section{Case Study}

To demonstrate the effectiveness of our results, we apply our approach to a room temperature regulation problem in a circular building. The model for this case study has been adapted from~\cite{Meyer.2018} by including stochasticity as a multiplicative noise. The evolution of the temperature $T(\cdot)$ in the interconnected system is governed by the following dynamics:
\begin{equation*}
\mathfrak{S}: 
T(k+1)=AT(k) + \mu T_H\nu(k) +  \theta T_E + 0.01\varsigma(k)T(k),
\end{equation*}
where $A \in \mathbb{R}^{n \times n}$ is a matrix with diagonal elements given by $\bar a_{ii}=(1-2\alpha-\theta-\mu\nu_i(k))$, off-diagonal elements $\bar a_{i,i+1}=\bar a_{i+1,i}=\bar a_{1,n}=\bar a_{n,1}= \alpha$, $i\in \{1,\ldots,n-1\}$, and all other elements are identically zero. The parameters  $\alpha=0.005$, $\theta=0.06$ and $\mu=0.145$ are conduction factors between rooms $i$ and $i \pm 1$, external environment and room $i$, heater and room $i$, respectively. The heater temperature is maintained at $T_H=40\,{}^{\circ}\mathsf{C}$ and  the outside temperature $T_{ei}=-5\,{}^{\circ}\mathsf{C}$ for all rooms $i \in \{1,\ldots,n\}$. We also have $T(k)=[T_1(k);\ldots ;T_n(k)]$, $T_E=[T_{e1};\ldots;T_{en}]$, $\nu(k)=[\nu_1(k);\ldots;\nu_n(k)]$ and $\varsigma(k)=\textsf{diag}(\varsigma_1(k),\ldots,\varsigma_n(k))$.

\begin{figure}
	\center
	\includegraphics[width=0.7\linewidth]{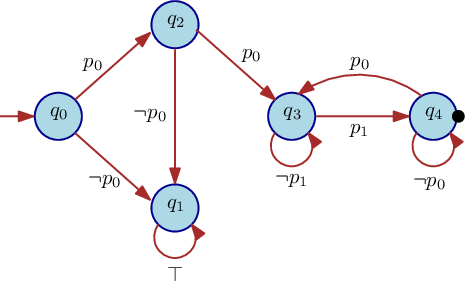}
	\caption{DSA $\mathcal{A}$ representing the specification with $\emph{Acc}=<q_4,\emptyset>$.}
	\label{temp_spec}
\end{figure}

The regions of interest are given by $X=[0,20]^n, X^{0}=[17,18]^n$, $X^{1}=[0,15]^n$, and $X^{2} = X \backslash (X^{0}\cup X^{1})$. 
We consider these regions to be associated with a set of atomic propositions $\mathcal{AP}=\{p_0,p_1,p_2\}$ via a labeling function $L: X \rightarrow \mathcal{AP}$ such that $L(x \in X^{z})=p_z$ for all $z \in \{0,1,2\}$.  The requirement of our case study is to synthesize a controller $\nu : \N \rightarrow [0,0.6]^n$ satisfying the specification represented by DSA $\mathcal{A}$ in Figure \ref{temp_spec} with $\emph{Acc}=<q_4,\emptyset>$. In order to achieve this, we must perform sequential decomposition on the DSA $\mathcal{A}$. To do this, we first obtain the set all lassos $\mathcal{R}^p$ for each $p \in \mathcal{AP}$. 
This can be obtained as $\mathcal{R}^{p_0}= \{(q_0,q_2,q_3,q_4,q_3,q_4) \}$. Correspondingly, the finite path set and the finite cycle set can be obtained as $\mathcal{R}^{p_0}_f=\{(q_0,q_2,q_3,q_4)\}$ and $\mathcal{R}^{p_0}_l=\{(q_4,q_3,q_4)\}$, respectively. As it can be seen, there is only one lasso $\tilde{\textbf{q}} \in \mathcal{R}^{p_0}$ which are decomposed into safety tasks. This is given by $\mathcal{P}^{p_0}(\tilde{\textbf{q}})= \{(q_0,q_2,q_3), (q_2,q_3,q_4), (q_3,q_4,q_3),(q_4,q_3,q_4)\}$. Furthermore, we have $\mathcal{P}^{p_0}(\tilde{\textbf{q}}_f)= \{(q_0,q_2,q_3), (q_2,q_3,q_4), (q_3,q_4,q_3)\}$ and $\mathcal{P}^{p_0}(\tilde{\textbf{q}}_l)=\{(q_4,q_3,q_4)\}$.  This constitutes four safety tasks for which we need to obtain CBCs and corresponding control policies. However, following Remark~\ref{rem:loopprob}, we prioritize the computation of CBC and corresponding control policy for the safety task $\vartheta=(q_4,q_3,q_4) \in \mathcal{P}^{p_0}(\tilde{\textbf{q}}_l)$.  

\begin{figure}
	\center
	\includegraphics[scale=0.8]{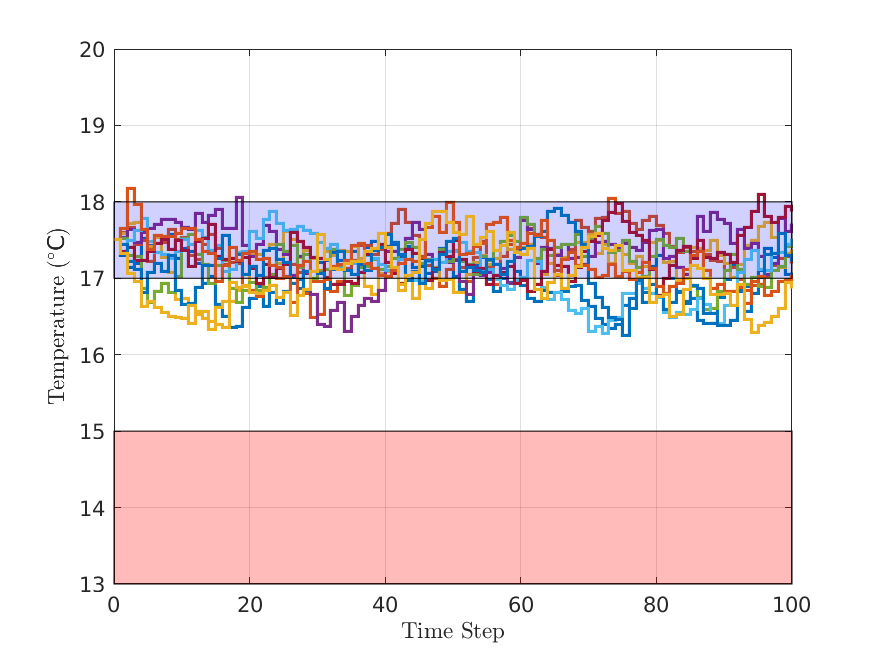}
	\caption{Closed-loop state trajectories of a representative room for $10$ noise realizations in a network of $300$ rooms, starting from a state in $X^2$. The region $X^0$ is shown in purple and $X^1$ is shown in pink.}
	\label{fig:simulation}
\end{figure}

To do this, we consider our network $\mathfrak{S}$ as an interconnection of $n=300$ subsystems, each of which constitutes a room. The state evolution of these individual subsystems is given by
\begin{align}\notag
\mathfrak{S}_i:
T_i(k+1) =  \bar aT_i(k) + \mu T_H \nu_i(k) +  \alpha w_i(k) + \theta T_{ei}  +  0.01\varsigma_i(k)T_i(k). \notag
\end{align}
It can be easily verified that $\mathfrak{S}=\mathcal{I}(\mathfrak{S}_1,\ldots,\mathfrak{S}_n)$ with coupling matrix $M$ such that $m_{i,i+1}=m_{i+1,i}=m_{1,n}=m_{n,1}=1$, $i \in \{1,\ldots,n-1\}$ and all other elements are identically zero. We now utilize the ADMM algorithm in conjunction with SOS formulation with the help of YALMIP tool \cite{Lofberg2004,Lofberg2009} to compute CSBCs for subsystems $\mathfrak{S}_i$, $i \in \{1,\ldots,N\}$.
Now, for the safety task $\vartheta$, we obtain CSBC as a $4^{th}$ order polynomial given by $\mathbb{B}_i(T_i)=9.6445-0.6911T_i+0.1396{T_i}^2-0.0163{T_i}^3+0.0005{T_i}^4$ and the corresponding controller is computed to be $\nu_i(T_i)=0.59-0.011T_i$.

Parameters satisfying conditions \eqref{subsys1}-\eqref{csbceq} are obtained as $\eta_i=0.0594$ and $\underbar X_i = 10^{-3} \times \begin{bmatrix}
0.1348 & 0.0001 \\ 0.0001 & -0.5591
\end{bmatrix}$. One can readily verify that compositionality conditions of Theorem \ref{Thm: Comp} are satisfied with $\eta=59.4$, $\beta=1000$ and $\underbar X^{comp}$ obtained from $\underbar X_i$, $i \in \{1,\ldots,300\}$, via equation \eqref{xcomp}. Therefore, the overall CBC of the interconnected system is obtained to be  $\mathbb{B}(T)=\sum_{i=1}^{300} (9.6445-0.6911T_i+0.1396{T_i}^2-0.0163{T_i}^3+0.0005{T_i}^4)$, while the suitable control policy for the trajectories in $X^{0}=L^{-1}(p_0)$ is obtained as $\nu_{p_0}(T)=[0.59-0.011T_1;\ldots;0.59-0.011T_{300}]$. The upper bound on the probability that the solution processes of the interconnected system $\mathfrak{S}$ start from $X_0=X^{0}$ and reach $X_u=X^{1}$ is computed to be equal to $0.0594$ by using Lemma \ref{lem:cbc}. 
However, from Theorem~\ref{thm:sumprod} and Remark~\ref{rem:loopprob},
	we can conclude that having the CSBC for $\vartheta \in \mathcal{P}^{p_0}(\tilde{\textbf{q}}_l)$ allows us to guarantee that the state $q_4$ is visited only finitely often with probability $1$, thereby allowing the satisfaction of DSA $\mathcal{A}$ with probability $1$. Therefore, it is not required to compute CBC for other safety tasks in $\mathcal{P}^{p_0}(\tilde{\textbf{q}})$, and instead we assign a pessimistic upper bound of $1$ for the violation of these safety tasks. The corresponding controllers are assumed to take any random value constrained within the input set. Finally, a switching mechanism for controllers is obtained as explained in Subsection \ref{control_policy}.

We now compute an overall probability of satisfaction of specification expressed by  DSA $\mathcal{A}$ when starting from an initial state $a \in X^{0}$ by using Theorem~\ref{thm:sumprod}:
\begin{equation*}
\mathbb{P}\{L(x^{a\nu}) \models \mathcal{A}\} = 1.
\end{equation*}  
Figure \ref{fig:simulation} shows the simulation for state trajectories of a representative room in the network for $10$ different noise realizations when starting from region $X^{2}$. The computation of CSBC and corresponding local control policy take up to $240$ seconds on a machine with Linux Ubuntu 18.04 OS (Intel i7-8665U CPU with 32GB RAM).  

\section{Discussion and Conclusion}
\subsection{Discussion}\label{subsec:disc}
The synthesis approach proposed in this work handles $\omega$-regular specifications described as DSA by decomposing them into a collection of safety properties. For these tasks, we construct control barrier certificates and corresponding control policies, as described in Section~\ref{subsec:seq_reach}. As suggested in Remark~\ref{rem:ignoreF}, this leads to ignoring the states in $F$ and considering only the states in $E$ of the DSA in order to perform sequential decomposition. Unfortunately, this is tailored to the nature of control barrier certificates which can only provide probabilistic guarantees over the satisfaction of \emph{safety} specifications. However, by reformulating suitable notions of CBCs for reachability specifications and combining it with the existing notions of CBCs for safety, one may be able to consider the states in $F$ and provide guarantees for visiting such states finitely often. Hence, one may be able to alleviate the underlying conservatism of our proposed approach. We leave further investigations in this direction to a future work.

\subsection{Conclusion}
In this paper, we proposed a compositional approach for the construction of control barrier certificates for large-scale interconnected discrete-time stochastic control systems. We first introduced notions of control sub-barrier certificates and control barrier certificates for control subsystems and interconnected systems, respectively. We then leveraged the interconnection topology to construct control barrier certificates of the interconnected system using control sub-barrier certificates of subsystems by imposing some dissipativity-type compositionality conditions. Obtained control barrier certificates were employed to provide probabilistic upper bounds for reaching unsafe regions. In order to extend this synthesis procedure for $\omega$-regular specifications, we also proposed a systematic approach to decompose specifications given by accepting languages of deterministic Streett automata into simple safety tasks. We then combined the probabilities obtained for the satisfaction of these tasks to obtain a lower bound on the probability of satisfaction of the original specification. We utilized alternating direction method of multipliers (ADMM) algorithm to obtain suitable control sub-barrier certificates for subsystems while ensuring satisfaction of compositionality conditions simultaneously. We also provided a sum-of-squares (SOS) formulation for systems with polynomial-type dynamics to compute appropriate polynomial-type control sub-barrier certificates along with their corresponding local control policies for subsystems. Finally, we illustrated our proposed methods by applying them to a room temperature regulation problem.

\section*{References}

\bibliographystyle{alpha}
\bibliography{nahs_arxiv}

\section{Appendix}

\begin{proof}{Theorem~\ref{probability_bounds}:}
	According to the condition~\eqref{sys3}, $X_u \subseteq \{x \in X \,|\, \mathbb{B}(x) \geq \beta\}$. Therefore, we have 
	\begin{align} \label{Eq:5}
	\mathbb{P}^{a}_{\varpi}\Big\{x(k)\in X_u \text{ for some } 0 \leq\, &k < \infty\,\,\big|\,\, a\Big\}  
	\notag \\ &\leq\mathbb{P}^{a}_{\varpi}\Big\{ \sup_{0\leq k < \infty} \!\!\mathds B(x(k))\geq \beta \,\big|\, a \Big\}.
	\end{align}
	Since the condition~\eqref{cbceq} implies that the CBC $\mathbb{B}$ is a non-negative supermartingale, the probability bounds ~\eqref{eq:bounds} follows directly from~\cite[Theorem 12, Chapter II]{1967stochastic} by applying it to ~\eqref{cbceq} and ~\eqref{sys2}.   
\end{proof}

\begin{proof} {Theorem \ref{Thm: Comp}:}
	First we show that the CBC $\mathbb{B}$ as \eqref{finalCBC} satisfies conditions \eqref{sys2} and \eqref{sys3}. For any $x\Let[x_{1};\ldots;x_{N}] \in X_0 = \prod_{i=1}^{N} X_{0_i} $ and from \eqref{subsys1}, we have
	\begin{align}\notag
	\mathds B(x) = \sum_{i=1}^N  \mathbb{B}_i(x_i)\leq  \sum_{i=1}^N  \eta_i = \eta,
	\end{align} 
	and similarly for any $x\Let[x_{1};\ldots;x_{N}] \in X_u = \prod_{i=1}^{N} X_{u_i} $ and from \eqref{subsys3}, one has
	\begin{align}\notag
	\mathds B(x) = \sum_{i=1}^N  \mathbb{B}_i(x_i)\geq N,
	\end{align} 	
	satisfying conditions \eqref{sys2} and \eqref{sys3} with $\eta=\sum_{i=1}^N  \eta_i$ and $\beta=N$. Since $\sum_{i=1}^N  \eta_i < N$ according to~\eqref{eq:eta}, one has $\beta > \eta$. Now we show that $\mathbb{B}(x)$ satisfies the condition \eqref{cbceq} as well. For any $i \in \{1,\ldots,N\}$, let there exist $\nu_i \in U_i$, with $\nu=[\nu_1;\ldots;\nu_N] \in U$ satisfying the condition \eqref{csbceq} and internal inputs given as $[w_1;\ldots;w_N]=M[h_{1}(x_1);\ldots;h_{N}(x_N)]$. Then, we can reach the chain of inequalities in \eqref{comproof} which completes the proof.
\end{proof}

\begin{figure*}[h!]
	\footnotesize
	\rule{\textwidth}{0.1pt}
	\begin{align} \notag
	\mathbb{E}\Big[&\mathds B(f(x,\nu,\varsigma)\,\big|\,x,{\nu}\Big] -\mathds B(x)=\mathbb{E}\Big[\sum_{i=1}^N\mathds B_{i}(f_i(x_i,\nu_i,w_i,\varsigma_i))\,\big|\,x,{\nu},w\Big]-\sum_{i=1}^N\mathds B_{i}(x_i)\\\notag
	&=\sum_{i=1}^N\mathbb{E}\Big[\mathds B_{i}(f_i(x_i,\nu_i,w_i,\varsigma_i))\,\big|\,x_i, \nu_{i},w_i\Big]-\sum_{i=1}^N\mathds B_{i}(x_i)\le \sum_{i=1}^N  \begin{bmatrix}
	w_i  \\ h(x_i)\end{bmatrix}^T \begin{bmatrix} \underbar X_i^{11} & \underbar X_i^{12} \\ 
	\underbar{X}_i^{21} & \underbar X_i^{22}  
	\end{bmatrix}\begin{bmatrix}
	w_i  \\ h(x_i)\end{bmatrix} \\\notag
	&\!=\!
	\begin{bmatrix}
	w_1 \\
	\vdots \\
	w_N \\
	h_{1}(x_1) \\
	\vdots \\
	h_{N}(x_N) 
	\end{bmatrix}^{\!T} \hspace{-0.5em} \begin{bmatrix}
	\underbar X_1^{11} &  &  &\underbar X_1^{12} &  &   \\ & \ddots &  &    & \ddots &  & \\
	&    &   \underbar X_N^{11}  &    &   &  \underbar X_N^{12} \\
	\underbar X_1^{21}  &   &   &  \underbar X_1^{22}  &   &   \\
	& \ddots &  &    & \ddots &  & \\
	&    &   \underbar X_N^{21}  &    &   &  \underbar X_N^{22} \\
	\end{bmatrix} \hspace{-0.5em} \begin{bmatrix}
	w_1 \\
	\vdots \\
	w_N \\
	h_{1}(x_1) \\
	\vdots \\
	h_{N}(x_N) 
	\end{bmatrix} \\
	& \!=\! \begin{bmatrix}
	M \begin{bmatrix} h_{1}(x_1) \\ \ldots \\ h_{N}(x_N) \end{bmatrix} \\
	h_{1}(x_1) \\
	\vdots \\
	h_{N}(x_N) 
	\end{bmatrix}^T \hspace{-0.7em}\underbar X^{comp} \hspace{-0.4em} \begin{bmatrix}
	M \begin{bmatrix} h_{1}(x_1) \\ \ldots \\ h_{N}(x_N) \end{bmatrix} \\
	h_{1}(x_1) \\
	\vdots \\
	h_{N}(x_N) 
	\end{bmatrix} \label{comproof}
	\!=\!    
	\begin{bmatrix}
	h_{1}(x_1) \\
	\vdots \\
	h_{N}(x_N)
	\end{bmatrix}^T \begin{bmatrix}
	M \\ I_{\tilde r} 
	\end{bmatrix}^T \underbar X^{comp} \begin{bmatrix}
	M \\ I_{\tilde r} 
	\end{bmatrix} \begin{bmatrix}
	h_{1}(x_1) \\
	\vdots \\
	h_{N}(x_N)
	\end{bmatrix} \le 0. 
	\end{align}
	\rule{\textwidth}{0.1pt}
\end{figure*}

\begin{proof} {Lemma \ref{sos}:}
	Since  $\lambda_{0_i}(x_i)$ in~\eqref{sos1} is sum-of-squares, we consequently have that the $\lambda_{0_i}^T(x_i)b_{0_i}(x_i) \geq 0$ in the region described by $X_{0_i} = \{x_i \in \R^{n_i} \mid b_{0_i}(x_i) \geq 0\}$. Since $\mathbb B_i(x_i)$ is also sum-of-squares and thus non-negative, condition~\eqref{sos1} directly implies the satisfaction of condition condition~\eqref{subsys1}.
	Similarly, we can show that \eqref{sos3} implies condition \eqref{subsys3}. Now, consider and \eqref{sos4}. If we choose the control input $\nu_{ji}=\lambda_{\nu_{ji}}(x_i)$, then since the terms $\lambda_i^T(x_i)b_i(x_i)$ and ${\lambda_i}^T_{w_i}b_{w_i}(w_i)$ are non-negative over $X$ and $W$, respectively, we can prove that it implies \eqref{csbceq}. This completes the proof. 
\end{proof}

\begin{proof}{Theorem \ref{thm:sumprod}:}Consider the set of lassos $\mathcal{R}^{p}$ and its corresponding set of finite paths and cycles $\mathcal{R}^p_f$ and $\mathcal{R}^p_l$ for all $p \in \mathcal{AP}$. Let the sets $\mathcal{P}^p(\tilde{\textbf{q}})$, $\mathcal{P}^p(\tilde{\textbf{q}}_f)$, and $\mathcal{P}^p(\tilde{\textbf{q}}_l)$ consist of all the safety tasks obtained from these sets, respectively. Following Remark~\ref{rem:ignoreF}, to compute lower bound on the probability of satisfaction of the specification expressed by DSA $\mathcal{A}$, it is sufficient to compute the lower bound on the probability that the states in $E$ are not visited infinitely often. This lower bound can then be computed by first computing the probability upper bound of the states in $E$ visiting infinitely often.

To do this, we consider any $\vartheta=(q,q',q'') \in \mathcal{P}^p(\tilde{\textbf{q}})$, and obtain from Lemma~\ref{lem:cbc} the upper bound on the probability that the solution process of dt-SCS $\mathfrak{S}$ starts from $X_0=L^{-1}(\sigma(q,q'))$ and reaches $X_u=L^{-1}(\sigma(q',q''))$ under the control input $\nu_{\vartheta}$. This is given by $\frac{\eta_\vartheta}{\beta_\vartheta}$. 

In order to compute the upper bound on the probability that the states in $E$ are visited infinitely often, one requires to compute the upper bound on the probability that the solution process follows the lasso $\bar{\textbf{q}}=(q_0^f,q_1^f,\ldots,q_{a_f}^f,(q_0^l,q_1^l,\ldots, \\ q_{a_l}^l)^{\omega})$ starting from $X_0=L^{-1}(\sigma(q_0^f,q_1^f))$, which consists of finite paths $\tilde{\textbf{q}}_f$ repeated once and the finite cycles $\tilde{\textbf{q}}_l$ repeated infinitely many times. This is obtained as 	
	\begin{equation*}
	\mathbb{P}\big\{L(x^{a\varpi}) \nvDash \mathcal{A}\big\} \leq \!\!	
	\prod_{\vartheta \in \mathcal{P}^p(\tilde{\textbf{q}})} \! \! 
	\begin{cases} 
	\frac{\eta_\vartheta}{\beta_\vartheta} \hspace{-0.5em} &  \vartheta\!\notin\! \mathcal{P}^p(\tilde{\textbf{q}}_l), \\
	0  & \vartheta\!\in\! \mathcal{P}^p(\tilde{\textbf{q}}_l) \text{ and }  \eta_\vartheta < \beta_\vartheta, \\
	1 &  \vartheta\!\in\! \mathcal{P}^p(\tilde{\textbf{q}}_l) \text{ and } \eta_\vartheta \nless \beta_\vartheta.
	\end{cases}
	\end{equation*}
	Now given the initial condition $a \in L^{-1}(p)$, the upper bound for a solution process $x^{a \varpi}$ of $\mathfrak{S}$ to satisfy the condition of visiting the states in $E$ infinitely many times is basically the summation of probabilities of all possible lassos in $\mathcal{R}^p$, and is obtained by
	\begin{equation*}
	\mathbb{P}\big\{L(x^{a\varpi}) \nvDash \mathcal{A}\big\} \leq \!\!	
	\sum_{\tilde{\textbf{q}} \in \mathcal{R}^p} \prod_{\vartheta \in \mathcal{P}^p(\tilde{\textbf{q}})} \! \! 
	\begin{cases} 
	\frac{\eta_\vartheta}{\beta_\vartheta} \hspace{-0.5em} &  \vartheta\!\notin\! \mathcal{P}^p(\tilde{\textbf{q}}_l), \\
	0  & \vartheta\!\in\! \mathcal{P}^p(\tilde{\textbf{q}}_l) \text{ and }  \eta_\vartheta < \beta_\vartheta, \\
	1 &  \vartheta\!\in\! \mathcal{P}^p(\tilde{\textbf{q}}_l) \text{ and } \eta_\vartheta \nless \beta_\vartheta.
	\end{cases}
	\end{equation*}
Having the probability upper bound for visiting the states in $E$ infinitely often, one obtains the lower bound on the probability of visiting the states in $E$ finitely often, or in other words, the satisfaction of the specification expressed by DSA $\mathcal{A}$ as in inequality~\eqref{eq:sumprod}. This completes the proof.

\end{proof}		

\end{document}